\documentclass[11pt]{article}

\usepackage{amsthm,amsmath,amsfonts,amssymb}
\usepackage{natbib}
\usepackage{hyperref}
\usepackage{graphicx}
\usepackage{epsfig,color,bm,url,hhline,subcaption}
\usepackage{enumerate}
\usepackage{rotating}
\allowdisplaybreaks
\usepackage{authblk}
\usepackage[justification=centering]{caption}
\usepackage{comment}
\usepackage{booktabs}
\usepackage{array}
\usepackage{setspace}
\usepackage{indentfirst}
\usepackage[title]{appendix}
\usepackage{multirow}
\usepackage{mathrsfs}
\usepackage{mathtools}
\usepackage{makecell}
\usepackage [english]{babel}
\usepackage [autostyle, english = american]{csquotes}
\MakeOuterQuote{"}
\usepackage{subfiles}
\usepackage{cleveref}

\numberwithin{equation}{section}

\theoremstyle{plain}

\newtheorem{remark}{{Remark}}

\theoremstyle{remark}

\newtheorem*{example}{Example}

\def\marginnote#1{\setbox0=\vtop{\hsize4pc
		\small\raggedright\noindent\baselineskip9pt \rightskip=0.5pc plus
		1.5pc #1}\leavevmode \vadjust{\dimen0=\dp0
		\kern-\ht0\hbox{\kern-4.00pc\box0}\kern-\dimen0}}
\def\lboxit#1{\vbox{\hrule\hbox{\vrule\kern6pt
			\vbox{\kern6pt#1\kern6pt}\kern6pt\vrule}\hrule}}

\usepackage{threeparttable}

\usepackage{xr}
\makeatletter
\newcommand*{\addFileDependency}[1]{
  \typeout{(#1)}
  \@addtofilelist{#1}
  \IfFileExists{#1}{}{\typeout{No file #1.}}
}
\makeatother

\begin{document}
\title{Statistical Learning of Trade Credit Insurance Network Data\\
with Applications to Ratemaking and Reserving}
\author[a]{Woongchae Yoo}
\author[b]{Spark C. Tseung}
\author[a]{Tsz Chai Fung}
\affil[a]{\footnotesize \emph{Maurice R. Greenberg School of Risk Science, Georgia State University}}
\affil[b]{\footnotesize \emph{Department of Statistical Sciences, University of Toronto}}
	\date{}
	\maketitle
\begin{abstract} 

Trade credit insurance (TCI) is a specialized line of property and casualty insurance, protecting businesses against financial losses due to buyer's insolvency. Predictive modeling for TCI claims poses formidable challenges due to the data’s complexity, yet remains underexplored in the literature. Leveraging six years of detailed TCI data from an Asian TCI insurer, we develop a bivariate, network-augmented Generalized Linear Mixed Model (GLMM) to jointly model claim probability and reporting time gaps. Our model integrates extended-order degree centrality and random effects at the business and policy levels, adjusted for data incompleteness, to capture claim histories, reporting time gaps, and network relationships specific to TCI data. To implement a feasible workaround for the high-dimensional integrations required by individual random effects, we propose a scalable Stochastic Expectation-Maximization (SEM) algorithm. Data analysis using this TCI dataset demonstrates that our model significantly outperforms benchmark models in both model fit and predictive accuracy, highlighting the effectiveness of our approach for improved ratemaking and reserving in TCI. Supplementary materials for this article are available as an online supplement.
\end{abstract}
\newpage

\section{Introduction} \label{sec:intro}

\begin{figure}[b]
\begin{center}
\includegraphics[width=4.5in]{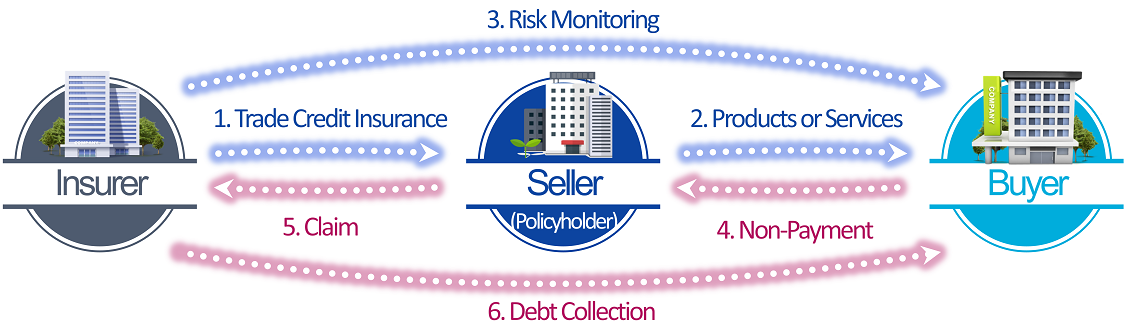}
\end{center}
\vspace{-1em}
\caption{Mechanisms of trade credit insurance operations.
\label{fig:Mechanisms of Trade Credit Insurance Operations}}
\end{figure}

Trade credit insurance (TCI) has emerged as a distinct business line of property and casualty (P\&C) insurance, serving to shield businesses (sellers) from potential losses in the event that their customers (buyers) become insolvent. As a widely used form of protection in commercial trade, TCI covered shipments valued at \$7 trillion globally in 2022, representing 13.16\% of worldwide trade in goods, according to the International Credit Insurance and Surety Association.
TCI plays a crucial role in safeguarding the financial health of businesses, particularly those engaging in transactions with open account terms which expose them to risks of insolvency and liquidity challenges (\cite{jones2010trade}). By mitigating systemic risk, TCI can serve as the last resort to prevent chain bankruptcies during economic crises. \Cref{fig:Mechanisms of Trade Credit Insurance Operations} illustrates the entire process of TCI. Given its significance, developing predictive models for ratemaking and reserving in TCI is vital to ensure fair premiums for policyholders and adequate capital set up for insurers.

Developing predictive models for TCI claims is a significant challenge due to the complex nature of TCI data. While sharing some characteristics with conventional P\&C insurance claim data, such as longitudinal data, where past claim history affects future claim predictions (see, e.g., \cite{shi2018pair} and \cite{yang2019multiperil}), and data truncation resulting from reporting time gaps, where claims reported after the evaluation date are not yet recorded (see, e.g., \cite{antonio2014micro} and \cite{verbelen2022modeling}), TCI claim data introduce an additional layer of complexity. This complexity arises from the presence of an intricate buyer-seller directional network structure, leading to more convoluted claim dependencies among and within policies compared to traditional P\&C claim data. Examples of dependencies from the network structure include:
\begin{itemize}
    \item \textbf{Policies sharing the same buyer:} A buyer may involve in multiple insurance policies. In the event of a buyer's default, multiple sellers can file claims if they have unpaid balances owed by the buyer. Therefore, the loss behavior across different policies sharing the same buyer can be correlated, known as a "contagion" effect.
    \item \textbf{Simultaneous buyer-seller dynamics:} Any buyer can also be a seller, and vice versa. Businesses may also insure against each other, causing an interdependence between them.
    \item \textbf{Multiple-buyer policy:} Within the same policy, a seller can insure multiple buyers. As the trade connections within the same policy share a common risk factor, i.e., the seller, the resulting losses may be correlated.
\end{itemize}

Despite its importance and inherent challenges, predictive modeling of TCI has received limited attention in the literature, primarily due to its small global market size and the scarcity of available data. Existing research on TCI mainly explores its economic impact rather than focusing on predictive analysis (e.g., \cite{yang2021trade}). Most studies concentrate on export credit insurance, emphasizing international trade (e.g., \cite{van2015private}), whereas our focus is on domestic trade (e.g., \cite{li2016trade}). There have been very few attempts to model and price TCI; notable exceptions include applications of the CreditRisk+ (CR+) model and its variations (\cite{passalacqua2006pricing}; \cite{giacomelli2021improved}; \cite{giacomelli2023parametric}). However, these studies do not incorporate the intricate network-type dependencies present in TCI data, nor do they utilize individual business-level risk characteristics to differentiate pricing for different policies. 
In contrast, statistical modeling of longitudinal data in other insurance applications has been extensively studied. Copula models (e.g., \cite{frees2008hierarchical}; \cite{diers2012dependence}; \cite{zhang2013predicting}; \cite{frees2016multivariate}), random effects models (e.g., \cite{boucher2006fixed}; \cite{pechon2018multivariate}; \cite{tseung2023improving}), and finite mixture models (e.g., \cite{tzougas2021multivariate}) are commonly employed in traditional longitudinal claim data to capture serial claim dependence and predict future claim distributions based on past claim information. However, the unique network structure of TCI data necessitates a more dedicated model that better aligns with its specific characteristics.

Statistical network modeling has primarily focused on social applications (e.g., \cite{holland1981exponential} and \cite{hunter2008goodness}) but is increasingly applied in other fields like finance (\cite{ahelegbey2016econometrics}). A widely used model is the Exponential Random Graph Model (ERGM), which estimates the probability of observing a network based on dyadic relationships between nodes, considering features like the number of connections and triangles. See, e.g., \cite{salter2012review} and \cite{loyal2020statistical} for a comprehensive review. However, applying ERGM to our TCI dataset presents challenges. Traditional models like ERGM predict the likelihood of link formation between nodes but do not predict claim probabilities given existing connections. Moreover, these models typically consider networks with nodes and edges, whereas the TCI data involve more complex structures due to multiple-buyer policies, resulting in outward-pointing stars where one node connects to multiple others simultaneously.

To assess node importance, the concept of degree centrality (DC) is introduced. First-order DC (FODC) evaluates the number of direct links to or from each node, which is simple and interpretable (\cite{golbeck2013analyzing}; \cite{golbeck2015introduction}). While FODC captures local influence, it doesn't account for a node's broader impact within the network.
Other measures like closeness centrality, betweenness centrality, eigenvector centrality, and the effective distance gravity model interaction score (\cite{shang2021identifying}) are applicable mainly to fully connected graphs, which is not the case with the TCI dataset.

In this study, we aim to develop a modeling framework for ratemaking and reserving within the context of TCI, accounting for the intricate dependencies and data truncation present in TCI data. To the best of our knowledge, this is the first study to apply data-driven statistical learning methods to TCI claims using a unique individual business-level real TCI dataset. Our key contributions are as follows. We introduce an expanded directed-network variant of the Generalized Linear Mixed Model (GLMM), which jointly models the claim occurrence probability and reporting time gaps while addressing the impact of data incompleteness resulting from reporting time gaps. The model incorporates various levels of observed information, including buyer, seller, policy, and trade connection details. It also incorporates unobserved information at the buyer, seller, and policy levels to capture unobserved risk characteristics and to model the serial dependence of claim probabilities over time at various levels. 
Furthermore, the model effectively considers the diverse network dependencies by introducing second-order DC (SODC) measures for directed TCI networks. This measure quantifies how the relative importance of each business entity within the directed network graph influences predictive claims. Specifically, we include both FODC and SODC measures in our models as additional covariates reflecting the seller's or buyer's importance in the network. By integrating higher-order DC measures alongside FODC, we overcome the ``locality'' limitations associated with FODC. Given the complexities of our proposed model, employing conventional parameter estimation methods presents substantial computational challenges. To address this, we derive a Stochastic Expectation-Maximization (SEM) algorithm, which efficiently calibrates our model and generates predictive claim distributions. 
The combined contributions of our study have several practical applications:
\begin{enumerate}
    \item \textbf{Ratemaking:} Our model outputs predictive claim probabilities for new or renewal trade connections, based not only on observed risk characteristics and claim histories but also on TCI network dynamics. This aids insurers in determining appropriate premiums or financial institutions, who provide \textit{factoring} (\cite{klapper2006role}), in offering specialized discount rates based on the perceived risk level.
    \item \textbf{Reserving:} The model also outputs the probability that a claim will eventually be reported in the future, given that no claims have been reported for an existing trade connection. It is crucial for insurers to set aside adequate capital for these unreported claims to prevent insolvency issues.
    \item \textbf{Systemic Risk Management:} By capturing the interdependence of claims among network connections, insurers can better understand and mitigate systemic risks similar to those in financial markets (\cite{eisenberg2001systemic}).
\end{enumerate}

While our empirical study focuses on TCI, the modeling framework is more general: it delivers edge-level risk modeling on a directed network with time-varying covariates and latent entity effects under incomplete observations. This modeling framework can potentially be useful for wider applications. Within insurance, closely related settings include cyber insurance (\cite{fahrenwaldt2018pricing}, \cite{xu2019cybersecurity}) and business interruption/supply-chain insurance (\cite{rose2016improving}). Beyond insurance, analogous edge-centric problems appear in, e.g., online transaction fraud detections (\cite{kodate2020detecting}), flight delay predictions (\cite{sadeek2025examining}), and infectious disease transmission risk management (\cite{simmering2015hospital}, \cite{chang2021mobility}). Further discussions on these broader areas are leveraged to Section \ref{sec:discuss}.

The paper is organized as follows. \Cref{sec:data} provides an overview of TCI data. \Cref{sec:method:math frame} establishes a mathematical framework for TCI network graph. In \Cref{sec:method}, we propose a modeling framework and estimation algorithm for the TCI data. \Cref{sec:data anal} performs a real data analysis. \Cref{sec:discuss} concludes. Additional details regarding the estimation procedures are presented in Appendix A. The analysis code and synthetic TCI dataset can be accessed on \hyperlink{https://github.com/tszchai/TCI}{https://github.com/tszchai/TCI}.

\section{Data Overview} \label{sec:data}
This paper analyzes a proprietary TCI claim data from a major Asian insurance company, covering domestic transactions from 2015 to 2020. Although official figures are not publicly available, the insurer in this dataset is the dominant carrier in the domestic line nationally when measured by annual exposure. This dataset provides an extensive view of TCI structure, with detailed information on entities (buyers and sellers), policy specifics, trade connections, and claims. The dataset contains three main categories of information: entity data, policy and trade connection data, and claim data.

Entity data includes unique identifiers for each entity, which may function as a buyer, a seller, or both. The data also contains time-varying variables that reflect the risk characteristics of entities, such as listing status, industry classification, time in business, and sales amount. In total, there are 129,915 unique entities in the dataset, with 93,663 unique buyers, 53,915 unique sellers, and 17,663 entities acting as both buyers and sellers. See \Cref{table:Data Summary} for the summary.

\begin{table}[!h]
\caption{Summary statistics of trade connection and claim counts by policy year.}
\centering
\resizebox{\columnwidth}{!}{%
\begin{tabular}{l c c c c c c c}
\hline
 & \multicolumn{7}{c}{\textbf{Policy Year}} \\
 & \textbf{2015} & \textbf{2016} & \textbf{2017} & \textbf{2018} & \textbf{2019} & \textbf{2020} & \textbf{Overall} \\ [0.5ex]
\hline
Number of Observations & 40,033 & 43,721 & 49,886 & 51,965 & 53,278 & 55,389 & 294,272\\
Number of Unique Sellers & 11,480 & 13,606 & 14,887 & 14,277 & 14,578 & 15,841 & 53,915 \\
Number of Unique Buyers & 25,593 & 26,903 & 29,977 & 31,867 & 33,068 & 34,544 & 93,663 \\
Number of Unique Businesses & 35,011 & 38,230 & 42,169 & 43,386 & 44,860 & 47,342 & 129,915 \\
Number of Businesses in the Intersection & 2,062 & 2,279 & 2,695 & 2,758 & 2,786 & 3,043 & 17,663 \\
Number of Claims & 920 & 962 & 1,210 & 1,296 & 1,181 & 1,148 & 6,717\\
Proportion of Policies with Non-Zero Claims & 5.99\% & 5.38\% & 5.94\% & 6.40\% & 5.45\% & 5.01\% & 5.66\%\\
\hline
\end{tabular}%
}
\label{table:Data Summary}
\end{table}

Policy and trade connection data provide insights into the insurance policies and the associated trade relationships. Each policy has a unique identifier and may be a single-buyer or multiple-buyer policy. In multiple-buyer policies, each trade connection between a buyer and a seller is given a unique trade connection identifier, resulting in multiple entries for the same policy number. All trade connections under the same policy share identical policy start and end dates. The data also maps each trade connection to its corresponding buyer and seller entities via unique identifiers, and contains policy-specific covariates (policy type, average turnover ratio, and policy insured amount) and trade connection covariates (buyer-specific insured amount and turnover ratio). The dataset records 104,494 unique policies and 294,272 unique insured trade connections.

Claim data links each claim to its corresponding trade connection via unique identifiers. It also includes the claim reporting date. Since a buyer’s default history prevents them from being insured again, a buyer can default only once.

Policies and trade connections are included in the dataset if the policy start date falls within the years 2015 to 2020. Buyer and seller information is included as long as the entity is associated with at least one active policy during this period. For the claim data, we continuously monitors claims associated with each observed policy reported up to June 30, 2023. Detailed descriptions of each variable are provided in \Cref{table:Description of Risk Characteristics}.

\begin{table}[!h]
\centering
\caption{Description and summary statistics of variables.}
\resizebox{\columnwidth}{!}{
\begin{tabular}{llccc}
\hline
\textbf{Category} & \textbf{Risk Characteristic} & \textbf{Range / Levels} & \textbf{Seller-Side} & \textbf{Buyer-Side} \\ \hline

\multicolumn{3}{l}{\textbf{Continuous Variables}} & \multicolumn{2}{c}{\textbf{Mean}} \\

Policy     & Total Insured Amount          & [1, 9800]           & \multicolumn{2}{c}{745.774}     \\
           & Average Turnover Ratio        & [2, 80]             & \multicolumn{2}{c}{6.162}       \\
Connection & Buyer-Specific Insured Amount & [1, 1000]           & \multicolumn{2}{c}{69.438}      \\
           & Buyer-Specific Turnover Ratio & [2, 189]            & \multicolumn{2}{c}{6.456}       \\
Entity     & Business Age                  & \begin{tabular}[c]{@{}c@{}}Seller: [0, 89]\\ Buyer: [0, 117]\end{tabular} & 11.709 & 15.503 \\

\hline
\multicolumn{3}{l}{\textbf{Categorical Variables}} & \multicolumn{2}{c}{\textbf{Proportion}} \\

Policy     & Policy Type                   & Single-Buyer           & \multicolumn{2}{c}{0.264}       \\
           &                               & Multiple-Buyer             & \multicolumn{2}{c}{0.736}       \\
Entity     & Business Type                 & Sole Proprietorship     & 0.135 & 0.043 \\
           &                               & Unspecified Corporation & 0.007 & 0.029 \\
           &                               & Limited Liability Company (LLC) & 0.737 & 0.645 \\
           &                               & Audit-Compliant Corporation (ACC) & 0.112 & 0.210 \\
           &                               & Listed                  & 0.009 & 0.074 \\
           & Industry                      & Manufacturing           & 0.472 & 0.499 \\
           &                               & Wholesale               & 0.458 & 0.277 \\
           &                               & Professional Services   & 0.028 & 0.031 \\
           &                               & Others                  & 0.042 & 0.193 \\
           & Annual Sales                  & Small [0, 5000]     & 0.343 & 0.268 \\
           &                               & Medium (5000, 20000]  & 0.395 & 0.269 \\
           &                               & Large (20000, $\infty$) & 0.203 & 0.363 \\
           &                               & Not Available           & 0.059 & 0.100 \\ \hline
\end{tabular}
}
\label{table:Description of Risk Characteristics}
\end{table}

All the data described above has been consolidated into a single aggregated dataset, such that each observation contains all buyer, seller, and policy-level characteristics, and claim information. \Cref{table:Example of Observation} provides an example of an observation.

\begin{table}[!h]
\caption{Example of an observation (trade connection) in the TCI dataset.}
\label{table:Example of Observation}
\resizebox{\columnwidth}{!}{%
\begin{tabular}{cccccccccc}
\hline
\begin{tabular}[c]{@{}c@{}}Policy\\      Number\end{tabular} & \begin{tabular}[c]{@{}c@{}}Policy\\      Type\end{tabular}           & \begin{tabular}[c]{@{}c@{}}Start\\      Date\end{tabular}            & \begin{tabular}[c]{@{}c@{}}Policy\\      Limit\end{tabular} & \begin{tabular}[c]{@{}c@{}}Avg Turnover\\       Ratio\end{tabular} & Seller                                                           & \begin{tabular}[c]{@{}c@{}}Seller\\      Business Type\end{tabular}    & \begin{tabular}[c]{@{}c@{}}Seller\\      Industry\end{tabular} & \begin{tabular}[c]{@{}c@{}}Seller\\      Business Age\end{tabular} & \begin{tabular}[c]{@{}c@{}}Seller\\      Annual Sales\end{tabular} \\ \hline
xxxxxx                                                                 & Multiple-buyer                                                                    & 5/19/2015                                                                     & 1,000                                                                                  & 4.72                                                                              & yyyyyy                                                                     & Sole Proprietorship                                                                 & Manufacturing                                                            & 11                                                                         & 47,252                                                                        \\ \hline
\multicolumn{1}{l}{}                                                  & \multicolumn{1}{l}{}                                                          & \multicolumn{1}{l}{}                                                          & \multicolumn{1}{l}{}                                                                & \multicolumn{1}{l}{}                                                              & \multicolumn{1}{l}{}                                                      & \multicolumn{1}{l}{}                                                      & \multicolumn{1}{l}{}                                                    & \multicolumn{1}{l}{}                                                       & \multicolumn{1}{l}{}            \\ \hline
Buyer                                                        & \begin{tabular}[c]{@{}c@{}}Insured\\      Amount\end{tabular} & \begin{tabular}[c]{@{}c@{}}Turnover\\      Ratio\end{tabular} & \begin{tabular}[c]{@{}c@{}}Buyer\\ Business Type\end{tabular}              & \begin{tabular}[c]{@{}c@{}}Buyer\\      Industry\end{tabular}            & \begin{tabular}[c]{@{}c@{}}Buyer\\      Business Age\end{tabular} & \begin{tabular}[c]{@{}c@{}}Buyer\\      Annual Sales\end{tabular} & Claim                                                          & \begin{tabular}[c]{@{}c@{}}Claim\\      Date\end{tabular}         & \begin{tabular}[c]{@{}c@{}}Claim\\      Amount\end{tabular}       \\ \hline
zzzzzz                                                                 & 100                                                                          & 5.24                                                                          & LLC                                                                                 & Wholesale                                                                         & 42                                                                        & 520,669                                                                      & Yes                                                                       & 9/27/2016                                                                  & 100    \\ \hline 
\end{tabular}%
}
\end{table}

Of the 294,272 insured trade connections observed, 238,883 from 2015 to 2019 are used for in-sample evaluation, with 80\% (191,107 observations) allocated for training and 20\% (47,776 observations) for validation. The remaining 55,389 observations from 2020 are set aside for out-of-sample testing. \Cref{table:Data Summary} presents summary statistics for claims observed up to June 30, 2023, showing that approximately 2.5\% of trade connections result in a claim, and 5.66\% of policies report at least one claim. The primary focus of this study is the claim probability, which will be applied to ratemaking and reserving.

\begin{figure}[!h]
\begin{center}
\includegraphics[width=0.6\textwidth]{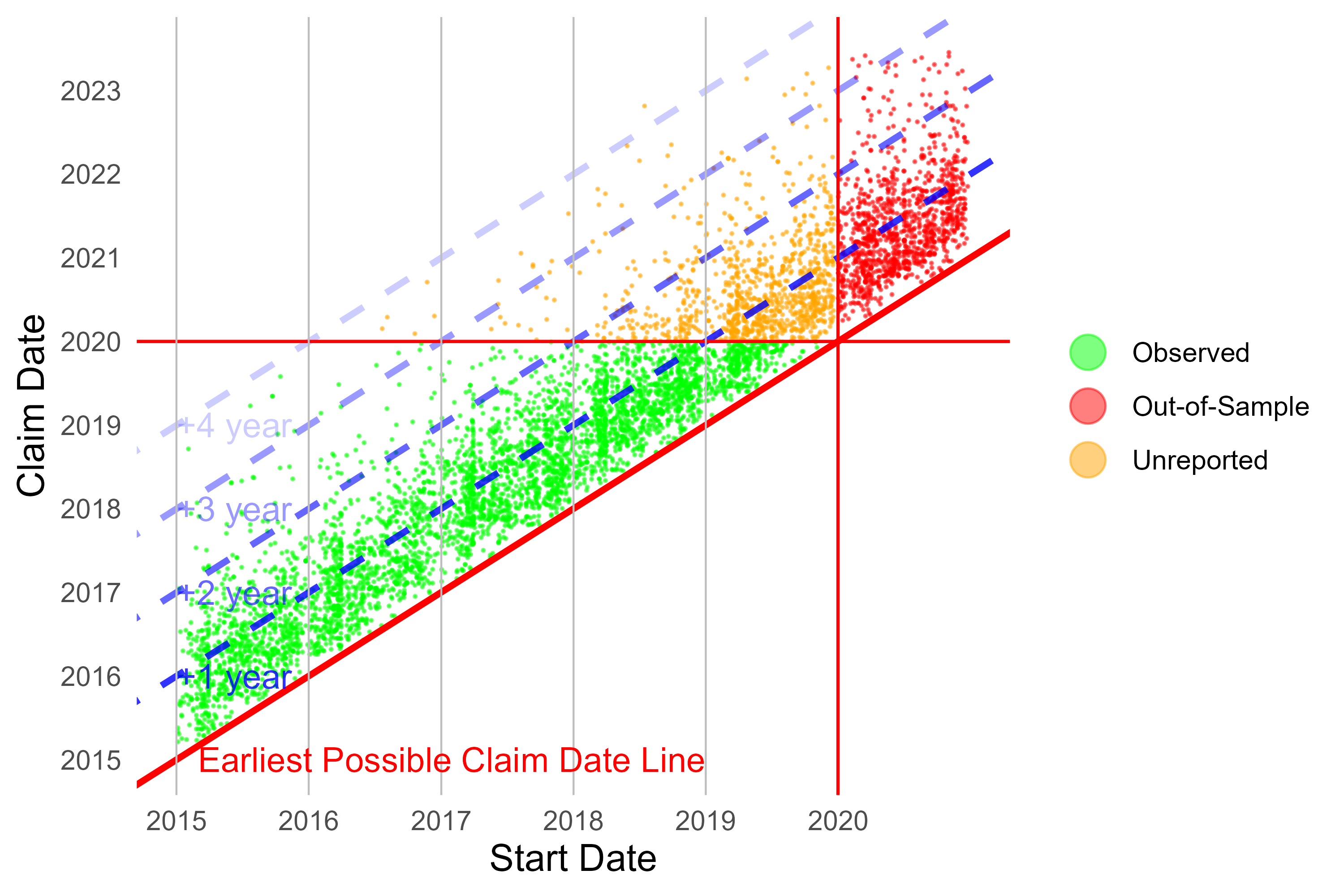}
\end{center}
\vspace{-1.5em}
\caption{Policy start date and claim date for each TCI claims. The vertical and horizontal red lines represent the evaluation date, i.e., end of the in-sample period. The diagonal red line shows the earliest possible claim date (policy start date), while the diagonal dotted lines mark successive years. Green, orange, and red dots are, respectively, claims observed by the evaluation date, unreported claims, and claims from out-of-sample policies.
\label{fig: Claims and Reporting Time Gaps}}
\end{figure}

In insurance practice, unlike our dataset where claim developments are tracked over an extended period, claims reported after the evaluation date are typically not observed, and reporting time gaps are common for TCI claims. Therefore, properly accounting for the effects of unobserved claims is essential to avoid underestimating claim probability. \Cref{fig: Claims and Reporting Time Gaps} illustrates the relationship between policy start dates and claim dates for TCI trade connections that result in a claim within our dataset. Most claims are reported within three years. However, a significant number of claims are reported after the evaluation date of December 31, 2019, making them unobserved in practice. Hence, it is crucial to adjust for the data incompleteness in our proposed model.

\begin{figure}[!h]
\begin{center}
\includegraphics[width=0.35\textwidth]{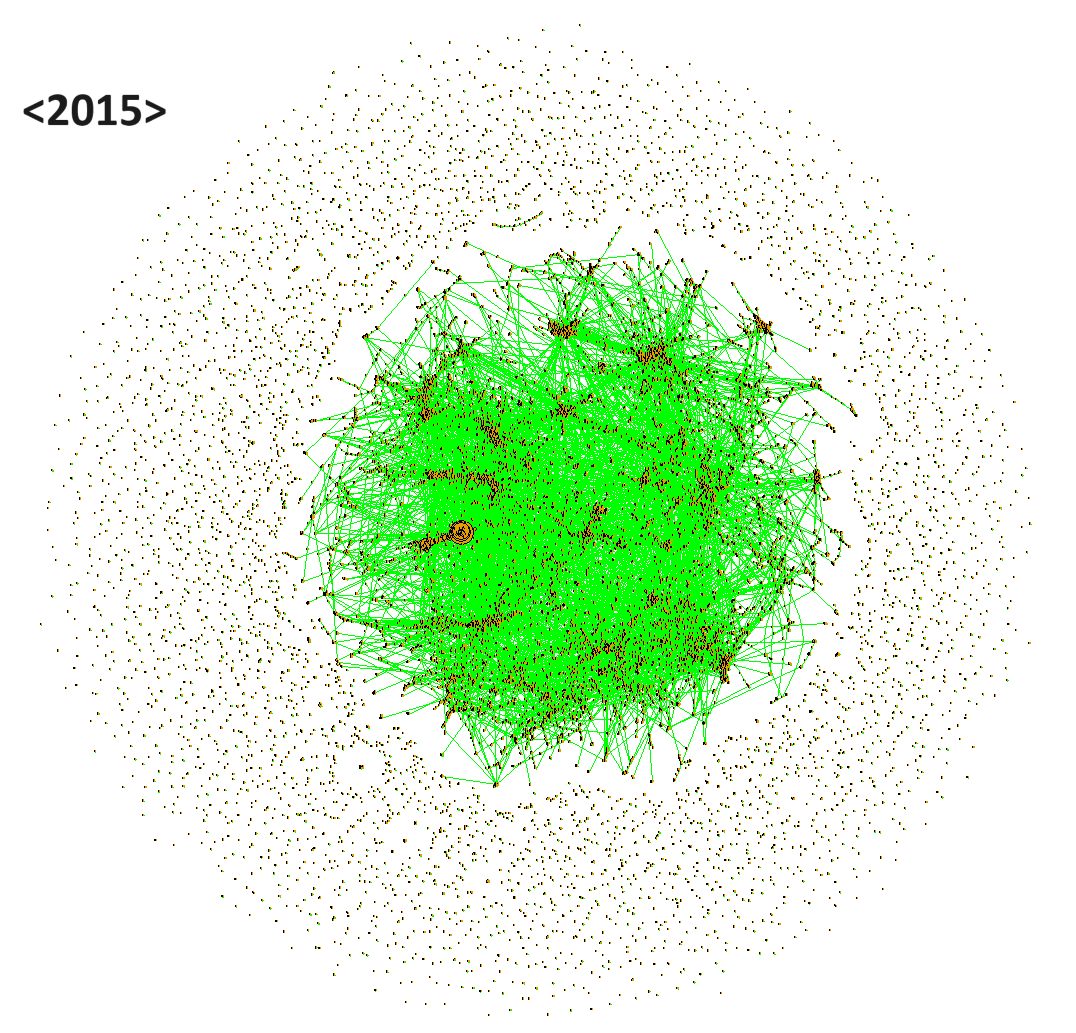}
\includegraphics[width=0.35\textwidth]{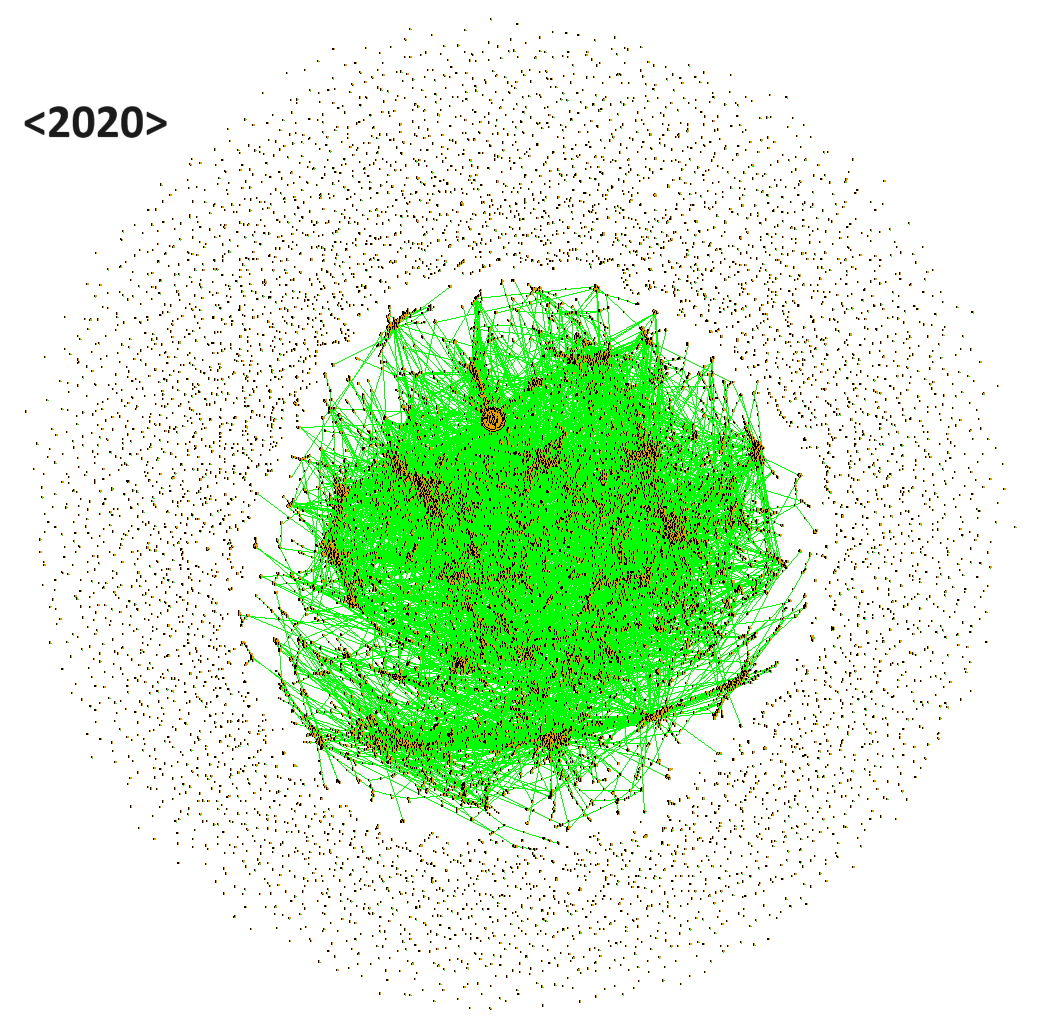}
\end{center}
\vspace{-1.5em}
\caption{Network graph for some policy years. Dots are entities, and green lines are trade connections.
\label{fig: Network Graph for Each Policy Year}}
\end{figure}

\Cref{fig: Network Graph for Each Policy Year} provides preliminary visualizations showing the relationships between sellers and buyers at the end of policy years 2015 and 2020. These sketches indicate the existence of networks within our dataset. For each year, a large, central network connecting most entities can be seen, along with several smaller, isolated networks. Also, we observe several entities with a high number of connections, suggesting that the connectivity and relative importance of entities should be taken into account when developing our model.

\section{Mathematical Framework for Network Graph} \label{sec:method:math frame}
The observed TCI data can be mathematically formulated as a network graph \(\mathcal{G}=(\mathcal{N},\mathcal{P},\mathcal{C})\), where \(\mathcal{N}\) is a set of nodes, \(\mathcal{P}\) is a set of (outward-pointing) stars, and \(\mathcal{C}\) is a set of directed edges. Each node \(i\in\mathcal{N}\) represents an involved entity, which can be a seller, a buyer, or both. Each star \(j\in\mathcal{P}\) represents a policy, where a seller is insured against the non-payment of one or more buyers. An edge \(k\in\mathcal{C}\) corresponds to a trade connection between a seller and a buyer that is insured, constituting a part of a policy.

\begin{figure}[!h]
\begin{center}
\includegraphics[width=0.49\textwidth]{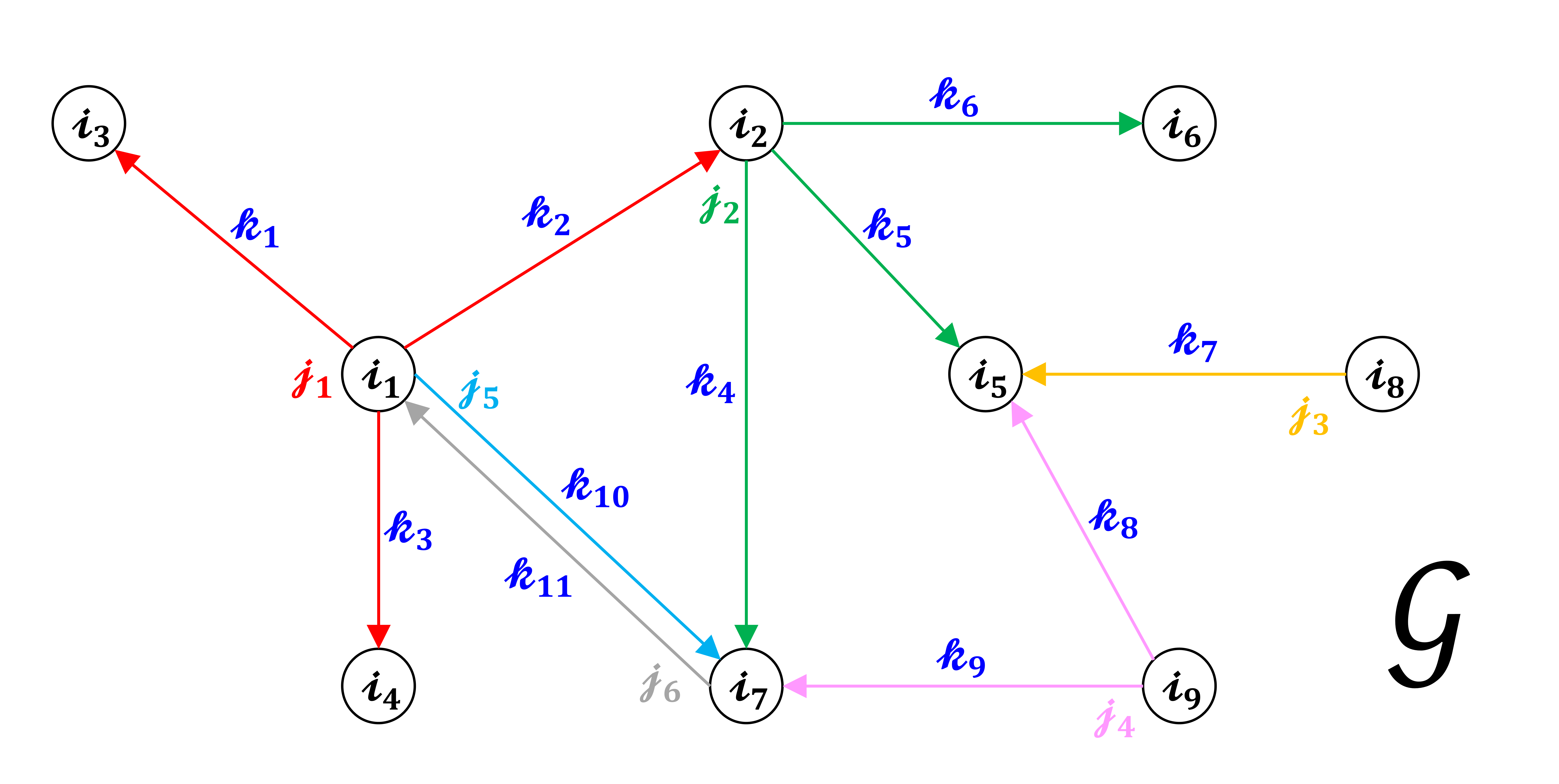}\hfill
\includegraphics[width=0.49\textwidth]{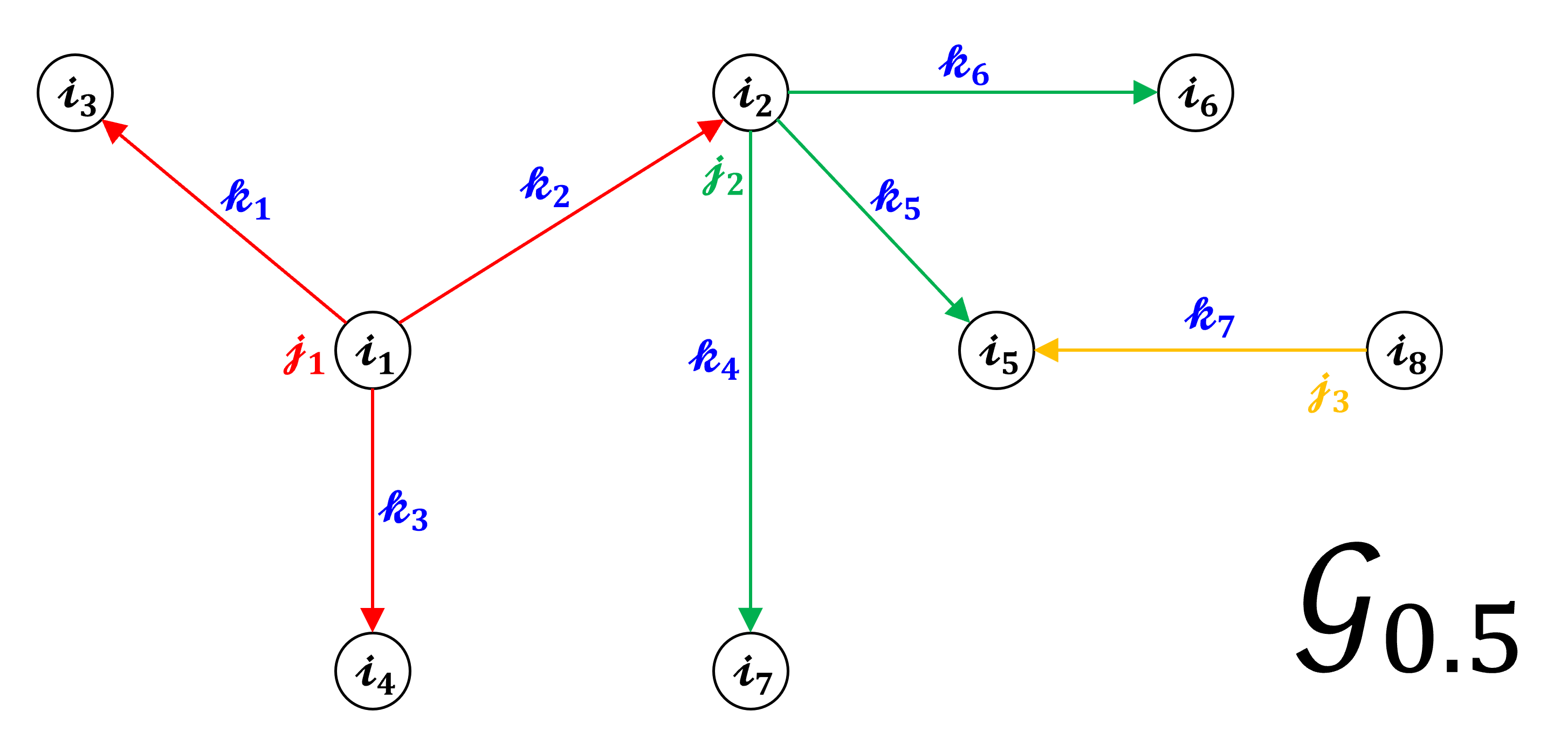}
\end{center}
\vspace{-1em}
\caption{Illustrative example of TCI network graph. Left panel: Full graph; Right panel: Sub-graph at $t=0.5$.
\label{fig:Example of Trade Credit Insurance Network Graph}}
\end{figure}

\begin{example}
The left panel of \Cref{fig:Example of Trade Credit Insurance Network Graph} illustrates a network structure of TCI. In this example, we have $\mathcal{N} = \{i_1, i_2, \ldots , i_9 \}$ with $|\mathcal{N}| = 9$ entities,  $\mathcal{P} = \{j_1, \ldots , j_6 \}$ with $|\mathcal{P}| = 6$ policies, and $\mathcal{C} = \{k_1, \ldots, k_{11} \}$ with $|\mathcal{C}| = 11$ trade connections.
\end{example}

We observe the function \(\mathscr{S}_P(\cdot): \mathcal{P} \rightarrow \mathcal{N}\), which maps each policy (or star) to its source (seller), and the function \(\mathscr{B}_P(\cdot): \mathcal{P} \rightarrow \sigma(\mathcal{N})\), mapping each policy to its targets (buyers), where \(\sigma(\mathcal{N})\) represents all possible subsets of \(\mathcal{N}\). Additionally, we observe the mappings \(\mathscr{S}_C(\cdot): \mathcal{C} \rightarrow \mathcal{N}\), \(\mathscr{B}_C(\cdot): \mathcal{C} \rightarrow \mathcal{N}\), and \(\mathscr{P}(\cdot): \mathcal{C} \rightarrow \mathcal{P}\), which associate each connection \(k \in \mathcal{C}\) respectively with the seller, buyer, and policy.

\begin{example}
Continuing from the previous example, we have, e.g., $\mathscr{S}_P(j_1) = i_1$, $\mathscr{B}_P(j_1) = \{i_2, i_3, i_4\}$, $\mathscr{B}_P(j_3) = \{i_5\}$, $\mathscr{S}_C(k_5) = i_2$, $\mathscr{B}_C(k_5) = \{i_5\}$, and $\mathscr{P}(k_5) = j_2$. Since $\mathscr{B}_P(j_1)$ contains more than one element, $j_1$ is a multiple-buyer policy. Conversely, $j_3$ is a single-buyer policy as $\mathscr{B}_P(j_3)$ only contains one element. Also, note that $\mathscr{S}_C(k_{10}) = \mathscr{B}_C(k_{11}) = i_1$ and $\mathscr{S}_C(k_{11}) = \mathscr{B}_C(k_{10}) = i_7$ depict a simultaneous buyer-seller case mentioned in \Cref{sec:intro}.
\end{example}

Each node \(i \in \mathcal{N}\) is associated with a set of node-specific features \(\bm{X}_i(t)\), which vary dynamically over time \(t \in [0, \tau]\) and provide information about an entity (e.g., business type, age, size, and industry) that may aid in predicting future claims. These features are unique to each node \(i\) and depend solely on time \(t\), not on whether the node is a seller or buyer. For instance, the (annual) sales amount for node \(i\) might fluctuate over time but is independent of the node’s role. Here, \(\tau\) represents the current or evaluation date of the insurance portfolio. For each star \(j \in \mathcal{P}\), we observe the policy start and end dates \((\underline{d}_j, \overline{d}_j)\) with \(\underline{d}_j \in [0, \tau)\), along with some policy-specific, time-independent information \(\bm{U}_j\) (e.g., total insured amounts, average turnover ratio, and policy type). Each edge \(k \in \mathcal{C}\) has associated time-independent features \(\bm{V}_k\) (e.g., insured amounts and turnover ratio specific to the connection or buyer), an actual claim occurrence indicator \(Z_k \in \{0, 1\}\), and an actual reporting time gap \(T_k \in [0, \infty]\). We set \(Z_k = 1\) if a claim eventually occurs, and \(Z_k = 0\) otherwise. The reporting time gap \(T_k\) refers to the interval between the policy start date \(\underline{d}_{\mathscr{P}(k)}\) and the time when a claim is reported to the insurer, with \(T_k = \infty\) if \(Z_k = 0\). 

Note the actual variables \(Z_k\) and \(T_k\) may not be fully and directly observed from the TCI dataset, because a claim that actually occurs before the evaluation date $\tau$ but is reported after $\tau$ remains unobserved or unrecorded. As such, we denote \(\tilde{Z}_k\) and \(\tilde{T}_k\) as the observed claim indicator and observed reporting time gap from the TCI data. Formally, we have the following relationship between $(Z_k,T_k)$ and $(\tilde{Z}_k,\tilde{T}_k)$:
\[
\tilde{Z}_k = Z_k \mathbf{1}\{T_k \leq \tau - \underline{d}_{\mathscr{P}(k)}\}
\quad\text{and}\quad
\tilde{T}_k =
\begin{cases}
T_k, & \text{if } T_k \leq \tau - \underline{d}_{\mathscr{P}(k)},\\
\infty, & \text{otherwise}.
\end{cases}
\]
Therefore, $(Z_k,T_k)$ and $(\tilde Z_k,\tilde T_k)$ are not identical in the presence of reporting delay. In survival-analysis, $(\tau - \underline{d}_{\mathscr{P}(k)})$ can be viewed as an ``administrative cut-off'', which is a right-truncation point for the observable event times.

\begin{example}
Consider our previous example with $\tau=2, (Z_{k_3}, T_{k_3})=(Z_{k_4}, T_{k_4})=(1, 1.9)$, $(Z_{k_5}, T_{k_5})=(0, \infty)$, $(\underline{d}_{j_{1}},\overline{d}_{j_{1}})=(0,1)$, and $(\underline{d}_{j_{2}},\overline{d}_{j_{2}})=(0.2,1.2)$. Since $\underline{d}_{\mathscr{P}(k_3)} + T_{k_3}=\underline{d}_{j_1} + T_{k_3}=0+1.9=1.9 < 2 = \tau $, the claim from $k_3$ is observed, and hence $(\tilde{Z}_{k_3}, \tilde{T}_{k_3})=(Z_{k_3}, T_{k_3})=(1, 1.9)$. However, since $\underline{d}_{\mathscr{P}(k_4)}+T_{k_4}=\underline{d}_{j_2}+T_{k_4}=0.2+1.9=2.1 > 2 = \tau $, the claim from $k_4$ has not yet observed, so $(\tilde{Z}_{k_4}, \tilde{T}_{k_4})=(0, \infty)$. Also, $(\tilde{Z}_{k_5}, \tilde{T}_{k_5})=(0, \infty)$. 
\end{example}

The network structure of the TCI data evolves over time, as not all policies are active at any given time \(t\). A policy \(j \in \mathcal{P}\) is considered active at time \(t\) if \(\underline{d}_j < t \leq \overline{d}_j\). Once a policy is no longer active, the associated insurance connections cease to exist. This leads us to define a network graph \(\mathcal{G}_t := (\mathcal{N}_t, \mathcal{P}_t, \mathcal{C}_t)\), containing only valid nodes, stars, and edges at time \(t\). Specifically, we have \(\mathcal{N}_t = \{i \in \mathcal{N} : i = \mathscr{S}_P(j)~\text{or}~i \in \mathscr{B}_P(j)~\text{for some}~j \in \mathcal{P}_t\}\), \(\mathcal{P}_t = \{j \in \mathcal{P} : \underline{d}_j < t \leq \overline{d}_j\}\), and \(\mathcal{C}_t = \{k \in \mathcal{C} : \mathscr{P}(k) \in \mathcal{P}_t\}\). Here, \(\mathcal{N}_t\) includes all entities involved in at least one policy active at time \(t\), \(\mathcal{P}_t\) is the set of active policies, and \(\mathcal{C}_t\) contains every seller-buyer connection associated with any active policy.

\begin{example}
In our example, consider $(\underline{d}_{j_{s}},\overline{d}_{j_{s}})=(0.2(s-1),0.2(s-1)+1)$ for $s=1,\ldots,6$, and $t=0.5$. Then, the active network subgraph at $t=0.5$ is given by the right panel of \Cref{fig:Example of Trade Credit Insurance Network Graph} with $\mathcal{N}_{0.5} = \{i_1,\ldots, i_8\}$, $\mathcal{P}_{0.5} = \{j_1, \ldots, j_3 \}$, and $\mathcal{C}_{0.5} = \{k_1, \ldots, k_7\}$.
\end{example}

\section{Methodology} \label{sec:method}
\subsection{Modeling Framework} \label{sec:method:model frame}
In this section, we jointly model the actual claim indicator \(Z_k\) and the actual reporting time gap \(T_k\) using a network-augmented bivariate GLMM. The joint specification is formulated for the actual variables $(Z_k,T_k)$, not for the observed pair $(\tilde{Z}_k,\tilde{T}_k)$ defined earlier. Because the TCI data provide only $(\tilde{Z}_k,\tilde{T}_k)$, which may differ from $(Z_k,T_k)$ due to reporting delay, inference requires a truncation-adjusted method to prevent estimation bias. We present the modeling details here and defer the inference construction to Section \ref{sec:method:inference}. Define \(\mathcal{D}^{\text{full}}\) as a set containing all explanatory variables and network connection information from the TCI data, i.e.,
\begin{equation*}
\mathcal{D}^{\text{full}}=\left(\left\{\bm{X}_i(t),\bm{U}_j,\bm{V}_k,\underline{d}_j,\overline{d}_j\right\}_{(i,j,k)\in\mathcal{G},t\in[0,\tau]},\mathscr{S}_P(\cdot),\mathscr{B}_P(\cdot),\mathscr{S}_C(\cdot),\mathscr{B}_C(\cdot),\mathscr{P}(\cdot)\right).
\end{equation*}

Define also a collection of latent variables \(\mathcal{D}^{\text{lat}}=\{B_i, S_i, P_j\}_{i\in\mathcal{N},j\in\mathcal{P}}\), where $B_i$ and $S_i$ are interpreted as the unobserved risk characteristics of entity \(i\) as a buyer and a seller, respectively, while \(P_j\) is the unobserved information associated with policy \(j\). We develop a joint model of \((Z_k,T_k)|\mathcal{D}^{\text{full}}\) based on the following assumptions:

\begin{itemize}
\item{\it (A1)} \(\{(Z_k,T_k)\}_{k\in\mathcal{C}}\) is a sequence of independent random vectors given \(\mathcal{D}^{\text{full}}\) and \(\mathcal{D}^{\text{lat}}\).
\item{\it (A2)} \((Z_k,T_k)|(\mathcal{D}^{\text{full}},\mathcal{D}^{\text{lat}})\) has the same distribution as \((Z_k,T_k)|(\mathcal{D}_k^{\text{obs}},\mathcal{D}_k^{\text{lat}})\) for all \(k\in\mathcal{C}\) with \(\mathcal{D}_k^{\text{obs}}:=(\bm{X}_k^B,\bm{X}_k^S,\tilde{\bm{U}}_k,\bm{V}_k)\) and \(\mathcal{D}_k^{\text{lat}}:=(\tilde{B}_k,\tilde{S}_k,\tilde{P}_k)\), where \(\bm{X}_k^B:=\bm{X}_{\mathscr{B}_C(k)}(\underline{d}_{\mathscr{P}(k)})\) and \(\bm{X}_k^S:=\bm{X}_{\mathscr{S}_C(k)}(\underline{d}_{\mathscr{P}(k)})\) are respectively the observed information of the buyer and seller associated with a connection \(k\) evaluated at the policy start date \(\underline{d}_{\mathscr{P}(k)}\), \(\tilde{\bm{U}}_k:=\bm{U}_{\mathscr{P}(k)}\) is the policy information of connection \(k\), \(\bm{V}_k\) is the connection information, and \(\tilde{B}_k:=B_{\mathscr{B}_C(k)}\), \(\tilde{S}_k:=S_{\mathscr{S}_C(k)}\) and \(\tilde{P}_k:=P_{\mathscr{P}(k)}\) are the buyer, seller and policy-level latent variables corresponding to a connection \(k\).
\item{\it (A3)} The latent variables \((B_i,S_i)\) and \(P_j\) are independent for any \(i\in\mathcal{N}\) and \(j\in\mathcal{P}\). Further, \(\{(B_i,S_i)\}_{i\in\mathcal{N}}\) are iid across \(i\in\mathcal{N}\), and \(\{P_j\}_{j\in\mathcal{P}}\) are iid across \(j\in\mathcal{P}\).
\end{itemize}

\(\it (A1)\) asserts that the dependence among different seller-buyer connections can be fully explained by all observed information and the latent variables. Note that \(\{(Z_k, T_k)\}_{k\in\mathcal{C}}\) are unconditionally dependent, where the network dependence is captured by the latent variables $\mathcal{D}^{\text{lat}}$. 
\(\it (A2)\) asserts that the joint distribution of \((Z_k, T_k)\) for any given connection \(k\) is determined only by the observed entity, policy, and connection information, and directly connected latent variables, evaluated at the policy start date. This assumption is reasonable when the features do not change a lot over time. Conversely, \(\it (A2)\) also implies that some information regarding the graphical structure of the data is implicitly lost when capturing the joint distribution of \((Z_k, T_k)\). We will explain how we mitigate this issue in \Cref{sec:method:feature} by incorporating some measures of network graphical structure into the node features. \(\it (A3)\) is a standard model assumption for mixed effects models. Note that $B_i$ and $S_i$ may be correlated. We emphasize that \((Z_k, T_k)\) are unconditionally dependent but become conditionally independent once incorporating random effects. 

To avoid any potential confusion between the two sets of latent variables $(B_i,S_i,P_j)$ and $(\tilde{B}_k,\tilde{S}_k,\tilde{P}_k)$ defined above, we now clarify their relationships:
\begin{itemize}
\item $B_i$, $S_i$ and $P_j$ are the entity or policy level latent variables.  Specifically, for each entity $i \in \mathcal{N}$, $(B_i,S_i)$ encodes its unobserved risk characteristics when acting as buyer and seller, respectively; and for each policy $j \in \mathcal{P}$, $P_j$ encodes the unobserved policy effect.
\item  $\tilde{B}_k$, $\tilde{S}_k$ and $\tilde{P}_k$ are the latent variables assigned to trade connection $k \in \mathcal{C}$. They map to the entity or policy level latent variables via, e.g., \(\tilde{B}_k:=B_{\mathscr{B}_C(k)}\).
\item While each $B_i$, $S_i$, and $P_j$ is unique across its index set, the variables $\tilde{B}_k$, $\tilde{S}_k$, and $\tilde{P}_k$ may recur for multiple connections sharing the same buyer entity $i$, seller entity $i$, or policy $j$, respectively. For example, for any two distinct connections $k$ and $k'$ with the same buyer index \(\mathscr{B}_C(k) = \mathscr{B}_C(k') = i\), we have \(\tilde{B}_k=\tilde{B}_{k'}=B_i\).
\end{itemize}
With these assumptions and notations, we model \(Z_k|(\mathcal{D}^{\text{full}},\mathcal{D}^{\text{lat}})\) through a logistic regression,
\begin{align} \label{eq:mod:z}
&Z_k|(\mathcal{D}^{\text{full}},\mathcal{D}^{\text{lat}})\overset{\mathcal{D}}{=}Z_k|(\mathcal{D}_k^{\text{obs}},\mathcal{D}_k^{\text{lat}})\overset{\text{ind}}{\sim}\text{Bernoulli}(p_k),\nonumber\\
&\log\frac{p_k}{1-p_k}=\alpha_0+\bm{\alpha}_1^{\top}\bm{X}_k^B+\bm{\alpha}_2^{\top}\bm{X}_k^S+\bm{\alpha}_3^{\top}\tilde{\bm{U}}_k+\bm{\alpha}_4^{\top}\bm{V}_k+\beta_1\tilde{B}_k+\beta_2\tilde{S}_k+\beta_3\tilde{P}_k,
\end{align}
and \(T_k|(\mathcal{D}^{\text{full}},\mathcal{D}^{\text{lat}},Z_k=1)\) through a Gamma regression,
\begin{align} \label{eq:mod:t}
&T_k|(\mathcal{D}^{\text{full}},\mathcal{D}^{\text{lat}},Z_k=1)\overset{\mathcal{D}}{=}T_k|(\mathcal{D}_k^{\text{obs}},\mathcal{D}_k^{\text{lat}},Z_k=1)\overset{\text{ind}}{\sim}\text{Gamma}(\mu_k,\psi),\nonumber\\
&\log\mu_k=\gamma_0+\bm{\gamma}_1^{\top}\bm{X}_k^B+\bm{\gamma}_2^{\top}\bm{X}_k^S+\bm{\gamma}_3^{\top}\tilde{\bm{U}}_k+\bm{\gamma}_4^{\top}\bm{V}_k+\nu_1\tilde{B}_k+\nu_2\tilde{S}_k+\nu_3\tilde{P}_k.
\end{align}

The latent variables $B_i$, $S_i$ and $P_j$ are naturally modeled by normal distributions with 
\begin{equation} \label{eq:re-prior-assumption}
\begin{aligned}
    \begin{pmatrix}
    B_i\\ S_i
    \end{pmatrix}\overset{\text{iid}}{\sim} N\left(\bm{0},
    \begin{pmatrix}
    1 & \rho\\
    \rho & 1
    \end{pmatrix}\right),\quad
    P_j\overset{\text{iid}}{\sim} N(0,1).
\end{aligned}
\end{equation}

To ensure identifiability, we restrict a unit variance for each latent variable as the coefficients \((\beta_1,\beta_2,\beta_3,\nu_1,\nu_2,\nu_3)\) in (\ref{eq:mod:z}) and (\ref{eq:mod:t}) already govern the magnitudes of random effects. Also, one of the random effect coefficients, say, \(\beta_1\), is required to be non-negative.

\begin{remark}
Parametric reporting time gap or delay models based on the Gamma distribution as in \eqref{eq:mod:t} are standard in micro-level P\&C reserving; see, for example, \cite{wuthrich2008stochastic} and \cite{antonio2014micro}. While other specifications (e.g., Weibull or lognormal) are also plausible and can be readily accommodated in our framework, the paper’s focus is on joint modeling that captures the TCI network dependence and accounts for data incompleteness. Therefore, we refrain from presenting an extensive catalog of alternative reporting time gap distributions, which would divert the paper’s focus.
\end{remark}
\begin{remark}
We do not include an explicit calendar-year regressor in the proposed model above because our TCI data has a short panel (five training years with the sixth year held out; see Section \ref{sec:data}). Treating year as a categorical factor leaves the hold-out year without a level in the training data, making forecasting ill-posed. Imposing a linear time trend for forecasting would require extrapolation to the hold-out year well beyond the observed time range, which is fragile and inflates uncertainty. Instead, we implicitly accommodate time-variation through the covariates: policy and entity features are updated over time, and the network-based covariates introduced in Section \ref{sec:method:feature} likewise evolve over time. With a materially longer data history, one could augment the specification with a calendar-year explanatory variable, but under the present panel length we find our current approach more reliable for forecasting.
\end{remark}

\subsection{Feature Engineering} \label{sec:method:feature}

\((A2)\) in \Cref{sec:method:model frame} implicitly limits the information set to the direct neighbors of a connection \(k\) when identifying the joint distribution of \((Z_k, T_k)\), which restricts our understanding of the broader network structure. To enhance the model's predictive power and gain deeper insights into how the network's graphical structure influences the claim distribution, we extract additional node features \(\bm{X}_i(t)\) based on each node’s dynamic connectivity and significance over time \(t \in [0, \tau]\). To assess local importance, we consider two types of first-order degree centrality (FODC), outward and inward degree centrality, which measure, respectively, the number of links extending from and into a node. Define
\begin{equation}
\begin{aligned}
    DC{_O}{^{(1)}}(i, t) = \sum_{k \in \mathscr{C}_t} \sum_{j \in \mathscr{P}_t} w_k I \left\{\mathscr{S}_P(j)=i \right\}
    = \sum_{k \in \mathscr{C}_t} w_k  I \left\{\mathscr{S}_C(k)=i \right\},
\end{aligned}
\label{eq:dc out}
\end{equation}
\begin{equation}
\begin{aligned}
    DC{_I}{^{(1)}}(i, t) = \sum_{k \in \mathscr{C}_t} w_k  I \left\{\mathscr{B}_C(k)=i \right\},
\end{aligned}
\label{eq:dc in}
\end{equation}
where \(w_k\) denotes the relative importance of connection \(k\). For simplicity, we assume \( w_k = 1 \) for all \( k \), implying that each connection is equally important. Alternatively, different weighting assumptions can be made; for instance, \(w_k\) could be inversely proportional to the number of buyers associated with the policy for connection \( k \) (i.e., \( w_k = |\mathscr{B}_P(\mathscr{P}(k))|^{-1} \)) or proportional to the insured amount. Clearly, \(\sum_{i \in \mathscr{N}_i} DC{_O}{^{(1)}}(i, t) = \sum_{i \in \mathscr{N}_i} DC{_I}{^{(1)}}(i, t) = |\mathscr{C}_t|\). Let \(\bm{D}(t) \in \mathbb{N}^{|\mathscr{C}_t| \times |\mathscr{C}_t|}\) represent the adjacency matrix for the network of all active entities at time \( t \), where the \((i_1, i_2)\)-th element, \(\bm{D}_{(i_1,i_2)}(t)\), is the number of directed edges pointing from node \( i_1 \) to \( i_2 \). Hence, \(DC{_O}{^{(1)}}(i, t)\) and \(DC{_I}{^{(1)}}(i, t)\) are, respectively, the \(i\)-th row and column sums of \(\bm{D}(t)\).

While FODC is useful for evaluating a node’s relative importance in the network, it is limited to measuring local significance. To address this limitation, we introduce second-order DC (SODC) measures, which capture broader significance and help identify key nodes with extended influence, such as a buyer insured by highly influential sellers with numerous trade connections, thereby representing a higher risk to the insurer. For a node \(i\) at time \(t\), there are four distinct SODC measures: outward-outward, inward-inward, inward-outward, and outward-inward, defined respectively as
\begin{equation}
\begin{aligned}
    DC{_{OO}}{^{(2)}}(i, t) = \sum_{\substack{k \neq k';~k, k' \in \mathscr{C}_t}} w_k w_{k'} I \left\{\mathscr{S}_C(k)=i, \mathscr{S}_C(k')=\mathscr{B}_C(k) \right\},
\end{aligned}
\label{eq:dc oo}
\end{equation}
\begin{equation}
\begin{aligned}
    DC{_{II}}{^{(2)}}(i, t) = \sum_{\substack{k \neq k';~k, k' \in \mathscr{C}_t}} w_k w_{k'} I \left\{\mathscr{B}_C(k)=i, \mathscr{B}_C(k')=\mathscr{S}_C(k) \right\},
\end{aligned}
\label{eq:dc ii}
\end{equation}
\begin{equation}
\begin{aligned}
    DC{_{IO}}{^{(2)}}(i, t) = \sum_{\substack{k \neq k';~k, k' \in \mathscr{C}_t}} w_k w_{k'} I \left\{\mathscr{B}_C(k)=i, \mathscr{S}_C(k')=\mathscr{S}_C(k) \right\},
\end{aligned}
\label{eq:dc io}
\end{equation}
\begin{equation}
\begin{aligned}
    DC{_{OI}}{^{(2)}}(i, t) = \sum_{\substack{k \neq k';~k, k' \in \mathscr{C}_t}} w_k w_{k'} I \left\{\mathscr{S}_C(k)=i, \mathscr{B}_C(k')=\mathscr{B}_C(k) \right\},
\end{aligned}
\label{eq:dc oi}
\end{equation}
where \(w_k\) and \(w_{k'}\) are both set to be one as in the FODC for simplicity in the rest of paper. \(DC{_{OO}}{^{(2)}}(i, t)\) and \(DC{_{II}}{^{(2)}}(i, t)\) are calculated as the \(i\)-th row and column sums, respectively, of \((\bm{D}(t)^2 - \text{diag}(\bm{D}(t)^2))\), while \(DC{_{OI}}{^{(2)}}(i, t)\) and \(DC{_{IO}}{^{(2)}}(i, t)\) are computed as the row sums of \((\bm{D}(t)\bm{D}(t)^{\top} - \text{diag}(\bm{D}(t)\bm{D}(t)^{\top}))\) and \((\bm{D}(t)^{\top}\bm{D}(t) - \text{diag}(\bm{D}(t)^{\top}\bm{D}(t)))\), respectively, where \(\top\) denotes matrix transposition and \(\text{diag}(\cdot)\) creates a diagonal matrix by retaining only the diagonal elements. The subtraction of \(\text{diag}(\bm{D}(t)^2)\) or \(\text{diag}(\bm{D}(t)\bm{D}(t)^{\top})\) is done to avoid including “round trip” cases where \(k = k'\) in (\ref{eq:dc oo}) -- (\ref{eq:dc oi}). Note that both FODC and SODC are time-evolving due to the dynamic network structure of the TCI data, as discussed in \Cref{sec:method:math frame}.

\begin{example}
Considering our previous example at $t=0.5$ (right panel of \Cref{fig:Example of Trade Credit Insurance Network Graph}), for node $i_2$, the first-order outdegree $DC{_{O}}{^{(1)}}(i_2, 0.5)$ is 3 through $k_4$, $k_5$ and $k_6$, while the second-order in-outdegree centrality $DC{_{IO}}{^{(2)}}(i_2, 0.5)$ is 2 through $(k_2,k_1)$ and $(k_2,k_3)$.
\end{example}

The introduction of FODC and SODC measures adds a total of six additional node-specific features \(\bm{X}_i(t)\). Hence, each logistic and Gamma regression in (\ref{eq:mod:z}) and (\ref{eq:mod:t}) includes 12 additional parameters: six for each of \(X_k^S\) and \(X_k^B\). \Cref{table:DC_stat} provides summary statistics for each DC measure by policy year in our TCI data. Overall, we observe an increase in most DC measures over time, indicating the growing popularity of TCI insurance and a resulting increase in network complexity and connectivity in recent years. Also, a drastic increase of seller-side outdegree from 2018 to 2019 reflects a system and policy change that removes certain limits on the number of buyers involved in multiple-buyer policies.

\begin{table}[!h]
\caption{Summary statistics of DC variables for each policy year. "Seller-Side DC" and "Buyer-Side DC" provide summaries of DC measures after filtering for entities that are sellers and buyers, respectively. }
\label{table:DC_stat}
\resizebox{\columnwidth}{!}{%
\begin{tabular}{ccccccccccccccc}
\hline
\multicolumn{1}{l}{}          &      & \multicolumn{6}{c}{\textbf{Seller-Side DC}} &  & \multicolumn{6}{c}{\textbf{Buyer-Side DC}} \\ \hline
\multicolumn{2}{c}{Policy Year}      & 2015  & 2016  & 2017  & 2018  & 2019 & 2020 &  & 2015  & 2016  & 2017  & 2018 & 2019 & 2020 \\ \hline
\multirow{2}{*}{\(DC{_O}{^{(1)}}\)}      & max  & 49    & 53    & 57    & 57    & 67   & 80   &  & 50    & 52    & 58    & 58   & 74   & 77   \\
                              & mean & 0.2   & 0.5   & 0.5   & 0.7   & 4.7  & 4.8  &  & 0.3   & 0.6   & 0.6   & 0.6  & 0.6  & 0.6  \\
\multirow{2}{*}{\(DC{_I}{^{(1)}}\)}       & max  & 17    & 19    & 27    & 20    & 25   & 23   &  & 68    & 84    & 84    & 95   & 92   & 95   \\
                              & mean & 0.3   & 0.5   & 0.6   & 0.6   & 0.6  & 0.6  &  & 1.5   & 3.2   & 3.2   & 3.4  & 3.4  & 3.1  \\
\multirow{2}{*}{\(DC{_{OO}}{^{(2)}}\)} & max  & 69    & 103   & 166   & 173   & 283  & 334  &  & 93    & 183   & 188   & 176  & 225  & 288  \\
                              & mean & 0.1   & 0.2   & 0.3   & 0.4   & 4.4  & 4.5  &  & 0.1   & 0.4   & 0.4   & 0.4  & 0.5  & 0.6  \\
\multirow{2}{*}{\(DC{_{II}}{^{(2)}}\)}   & max  & 21    & 20    & 22    & 23    & 39   & 23   &  & 68    & 91    & 94    & 87   & 86   & 81   \\
                              & mean & 0.1   & 0.3   & 0.4   & 0.4   & 0.6  & 0.6  &  & 0.8   & 2.7   & 2.6   & 2.7  & 2.9  & 2.6  \\
\multirow{2}{*}{\(DC{_{OI}}{^{(2)}}\)}  & max  & 269   & 369   & 395   & 562   & 749  & 690  &  & 467   & 391   & 368   & 412  & 753  & 714  \\
                              & mean & 0.5   & 1.6   & 1.5   & 1.8   & 17.5 & 15.0 &  & 0.9   & 2.7   & 2.7   & 2.7  & 2.8  & 2.5  \\
\multirow{2}{*}{\(DC{_{IO}}{^{(2)}}\)}  & max  & 351   & 325   & 369   & 402   & 358  & 384  &  & 555   & 664   & 723   & 653  & 774  & 865  \\
                              & mean & 2.8   & 5.5   & 5.5   & 6.1   & 7.9  & 9.5  &  & 14.3  & 30.6  & 29.2  & 29.6 & 35.3 & 35.5 \\ \hline
\end{tabular}%
}
\end{table}

\subsection{Inference} \label{sec:method:inference}

This section develops the estimation algorithm under the modeling framework in Section \ref{sec:method:model frame}. Since the joint model is formulated for the actual pair $(Z_k,T_k)$, while the observed TCI data provide $(\tilde Z_k,\tilde T_k)$, treating $(\tilde Z_k,\tilde T_k)$ as if they were $(Z_k,T_k)$ would lead to biased estimation, typically underestimating claim probabilities. Therefore, the observed data likelihood must be built from the joint distribution of $(\tilde Z_k,\tilde T_k)$ implied by the joint model for $(Z_k,T_k)$ and the right-truncation rule. Specifically, one can show that $\tilde{Z}_k$ conditionally follows Bernoulli with probability $p^*_k := p_k F(\tau - \underline{d}_{\mathscr{P}(k)}; \mu_k, \psi)$, while $\tilde{T}_k$ given $\tilde{Z}_k=1$ conditionally follows a right-truncated Gamma distribution with density $f^*(\tilde{t}; \mu_k, \psi) := f(\tilde{t}; \mu_k, \psi)/F(\tau - \underline{d}_{\mathscr{P}(k)}; \mu_k, \psi)$ for $\tilde{t}\in[0,\tau - \underline{d}_{\mathscr{P}(k)}]$.

Let \(\bm{\Psi}=(\bm{\alpha}, \bm{\beta}, \bm{\gamma},\bm{\nu},\psi,\rho)\) denote the set of all model parameters in \eqref{eq:mod:z} to \eqref{eq:re-prior-assumption}, where \(\bm{\alpha}:=(\alpha_0,\bm{\alpha}_1^{\top},\ldots,\bm{\alpha}_4^{\top})^{\top}\), \(\bm{\beta}:=(\beta_1,\beta_2,\beta_3)^{\top}\), \(\bm{\gamma}:=(\gamma_0,\bm{\gamma}_1^{\top},\ldots,\bm{\gamma}_4^{\top})^{\top}\), and \(\bm{\nu}:=(\nu_1,\nu_2,\nu_3)^{\top}\). Denote \(\bm{Z}=\{Z_k\}_{k\in\mathcal{C}}\), \(\bm{\tilde{Z}}=\{\tilde{Z}_k\}_{k\in\mathcal{C}}\), \(\bm{T}=\{T_k\}_{k\in\mathcal{C}}\) and \(\bm{\tilde{T}}=\{\tilde{T}_k\}_{k\in\mathcal{C}}\) as vectors of actual and observed claim occurrence indicators and reporting time gaps across all trade connections. The observed data likelihood given the random effects is
\begin{align} \label{eq:infer:lik_obs_cond}
&\mathcal{L}^{\text{obs}}(\bm{\Psi};\bm{\tilde{Z}},\bm{\tilde{T}},\mathcal{D}^{\text{full}}|\mathcal{D}^{\text{lat}})
:=\prod_{k\in\mathcal{C}}\mathcal{L}_k^{\text{obs}}(\bm{\Psi};\tilde{Z}_k,\tilde{T}_k,\mathcal{D}_k^{\text{obs}}|\mathcal{D}^{\text{lat}}_k)=\prod_{k\in\mathcal{C}}\left[p^*_kf^*(\tilde{T}_k;\mu_k,\psi)\right]^{\tilde{Z}_k}\left(1-p^*_k\right)^{1-\tilde{Z}_k},
\end{align}
and the unconditional observed likelihood is
\begin{equation} \label{eq:infer:lik_obs}
\mathcal{L}^{\text{obs}}(\bm{\Psi};\bm{\tilde{Z}},\bm{\tilde{T}},\mathcal{D}^{\text{full}})
=\int_{\Omega}\mathcal{L}^{\text{obs}}(\bm{\Psi};\bm{\tilde{Z}},\bm{\tilde{T}},\mathcal{D}^{\text{full}}|\mathcal{D}^{\text{lat}})\prod_{i\in\mathcal{N}}\phi(B_i,S_i;\rho)\prod_{j\in\mathcal{P}}\phi(P_j)d\mathcal{D}^{\text{lat}},
\end{equation}
where \(\phi(\cdot,\cdot;\rho)\) and \(\phi(\cdot)\) are respectively the density functions of bivariate and univariate standard normal distributions, and \(\Omega=\mathbb{R}^{(2|\mathcal{N}|+|\mathcal{P}|)}\) is the space of \(\mathcal{D}^{\text{lat}}\). Note that the construction of \eqref{eq:infer:lik_obs_cond} involves a standard likelihood-based adjustment technique for truncation in survival analysis; see, e.g., Chapter 3.5 of \cite{klein2003survival}. Such a truncation-adjusted inference method for reporting delay has also been explored in the actuarial literature on P\&C reserving (e.g., \cite{badescu2019marked}; \cite{fung2022fitting}). Those papers, however, consider a simpler setting in which the claim-arrival process and the reporting delay are assumed independent, whereas in our paper the dependence between $Z_k$ and $T_k$ is governed by latent variables.

As \eqref{eq:infer:lik_obs} involves high-dimensional integration of non-standard function \eqref{eq:infer:lik_obs_cond}, it is computationally prohibitive to directly optimize \eqref{eq:infer:lik_obs}. Common non-stochastic approximations are also not feasible here. Gaussian-quadrature approximations are accurate only in low dimensions and their node count grows exponentially with the dimension of $\Omega$. In our case, $\dim(\Omega)=2|\mathcal{N}|+|\mathcal{P}|=364,324$, and because the latent effects are globally coupled, i.e., \eqref{eq:infer:lik_obs} cannot be factorized into lower-dimensional components, such an approximation method is not viable. Laplace-type approximations involve locating the joint mode and evaluating the associated curvature, which requires solving linear systems of size $2|\mathcal{N}|+|\mathcal{P}|$. The inversion of coefficient matrix itself already incurs $\mathcal{O}((2|\mathcal{N}|+|\mathcal{P}|)^3)$ time and $\mathcal{O}((2|\mathcal{N}|+|\mathcal{P}|)^2)$ memory, which is prohibitive at our scale. On the other hand, the complete data log-likelihood
\begin{align} \label{eq:infer:lik_com}
\ell^{\text{com}}(\bm{\Psi};\bm{Z},\bm{T},\mathcal{D}^{\text{full}},\mathcal{D}^{\text{lat}})
=&\sum_{k\in\mathcal{C}}\left[Z_k\log p_k+(1-Z_k)\log(1-p_k)+Z_k\log f(T_k;\mu_k,\psi)\right]\nonumber\\
&+\sum_{i\in\mathcal{N}}\log\phi(B_i,S_i;\rho)+\sum_{j\in\mathcal{P}}\log\phi(P_j),
\end{align}
is computationally feasible. For ease of implementation and scalability to large datasets, we propose a Stochastic Expectation-Maximization (SEM) algorithm (\cite{celeux1985sem}) for efficient estimation of parameters, avoiding the need to perform any high-dimensional integrations. Our algorithm iterates between the following steps until convergence.

\textbf{SE-Step}: In the \(t\)-th iteration, we compute the expectation of complete data log-likelihood \(\mathbb{E}\left[\ell^{\text{com}}(\bm{\Psi};\bm{Z},\bm{T},\mathcal{D}^{\text{full}},\mathcal{D}^{\text{lat}})|\bm{\tilde{Z}},\bm{\tilde{T}},\mathcal{D}^{\text{full}},\bm{\Psi}^{(t-1)}\right]\). However, direct computation, either analytically or with the use of naive Monte Carlo (MC) method, is challenging as the posterior distribution of \(\mathcal{D}^{\text{lat}}\) given \((\bm{\tilde{Z}},\bm{\tilde{T}})\) cannot be expressed analytically. Several standard MC approaches are also not viable in our case. Importance sampling draws from a proposal and reweights to approximate the posterior expectation. For high-dimensional, sharply concentrated posteriors, natural proposals often suffer from weight degeneracy and collapse in effective sample size. Rejection sampling accepts a proposal draw with probability proportional to the ratio of posterior to proposal. However, in high dimension, it is extraordinarily difficult to find an analytical proposal that uniformly dominates the posterior, and even if such an envelope exists the acceptance probability typically decays exponentially with dimension. Classical Gibbs sampling, which cycles through full conditional distributions of the latent variables, is also not applicable because the full conditionals induced by our nonconjugate likelihood and cross-connection coupling are not of any standard parametric form.

As such, we employ the Markov chain Monte Carlo (MCMC) method with Metropolis-Hastings (MH) algorithm to sample the posterior random effects $(B_{\mathscr{B}_C(k)}^{(m,t)},S_{\mathscr{S}_C(k)}^{(m,t)},P_{\mathscr{P}(k)}^{(m,t)})$ for $m=1,\ldots,M$, where $B_{\mathscr{B}_C(k)}^{(m,t)}$, $S_{\mathscr{S}_C(k)}^{(m,t)}$ and $P_{\mathscr{P}(k)}^{(m,t)}$ are the simulated posterior buyer, seller and policy-level random effects at the $m$-th (sub-)iteration. The details regarding the sampling procedures are leveraged to Appendix A.1. The proposed sampling procedures enable parallel computing for all levels of random effects, ensuring efficiency. Denote \(\mathcal{M}\subset\{1,\ldots,M\}\) as a set containing the indexes of MCMC samples above that we choose to retain. Also, denote \(\bar{\bm{X}}_k^{(m,t)}=(1,\bm{X}^{B\top}_k,\bm{X}^{S\top}_k, \tilde{\bm{U}}_k^{\top},\bm{V}_k^{\top},B_{\mathscr{B}_C(k)}^{(m,t)},S_{\mathscr{S}_C(k)}^{(m,t)},P_{\mathscr{P}(k)}^{(m,t)})^{\top}\). We simulate posterior samples of time gap \(T_k^{(m,t)}\) given \((\tilde{Z}_k,\tilde{T}_k,B^{(m,t)}_{\mathscr{B}_C(k)},S^{(m,t)}_{\mathscr{S}_C(k)},P^{(m,t)}_{\mathscr{P}(k)})\) for \(k\in\mathcal{C}\) and \(m\in\mathcal{M}\) using the following steps:
\begin{enumerate}
\item If \(\tilde{Z}_k=1\), set \(T_k^{(m,t)}=\tilde{T}_k\).
\item If \(\tilde{Z}_k=0\), simulate \(T_k^{(m,t)}\) from a random variable with density \(f(t;\mu_k^{(m,t)},\psi^{(t-1)})1\{t>\tau-\underline{d}_{\mathscr{P}(k)}\}/[1-F(\tau-\underline{d}_{\mathscr{P}(k)};\mu_k^{(m,t)},\psi^{(t-1)})]\), i.e., a left-truncated gamma distribution, where \(\log \mu_k^{(m,t)}=(\bm{\gamma}^{(t-1)\top},\bm{\nu}^{(t-1)\top})\bar{\bm{X}}_k^{(m,t)}\).
\end{enumerate}

Finally, the E-step requires the evaluation of
\begin{align}
Z_k^{(t)}
&:=\mathbb{E}[Z_k|\tilde{Z}_k,\mathcal{D}_k^{\text{obs}},B^{(m,t)}_{\mathscr{B}_C(k)},S^{(m,t)}_{\mathscr{S}_C(k)},P^{(m,t)}_{\mathscr{P}(k)}]\nonumber\\
&=\frac{p_k^{(m,t)}[1-F(\tau-\underline{d}_{\mathscr{P}(k)};\mu_k^{(m,t)},\psi^{(t-1)})]+(1-p_k^{(m,t)})\tilde{Z}_k}{p_k^{(m,t)}[1-F(\tau-\underline{d}_{\mathscr{P}(k)};\mu_k^{(m,t)},\psi^{(t-1)})]+(1-p_k^{(m,t)})},
\end{align}
where \(\log \left[p_k^{(m,t)}/(1-p_k^{(m,t)})\right]:=(\bm{\alpha}^{(t-1)\top},\bm{\beta}^{(t-1)\top})\bar{\bm{X}}_k^{(m,t)}\).

\textbf{M-Step}: In the \(t\)-th iteration, we maximize the following \(Q\)-function w.r.t. \(\bm{\Psi}\):
\begin{align} \label{eq:infer:Q}
Q(\bm{\Psi};\mathcal{D}^{\text{full}},\bm{\Psi}^{(t-1)})
&=\frac{1}{|\mathcal{M}|}\sum_{m\in\mathcal M}\sum_{k\in\mathcal{C}}\left[Z_k^{(m,t)}\log p_k^{(m,t)}(\bm{\alpha},\bm{\beta})+(1-Z_k^{(m,t)})\log (1-p_k^{(m,t)}(\bm{\alpha},\bm{\beta}))\right]\nonumber\\
&\quad +\frac{1}{|\mathcal{M}|}\sum_{m\in\mathcal M}\sum_{k\in\mathcal{C}}Z_k^{(m,t)}f(T_{k}^{(m,t)};\mu_k^{(m,t)}(\bm{\gamma},\bm{\nu}),\psi)\nonumber\\
&\quad +\frac{1}{|\mathcal{M}|}\sum_{m\in\mathcal M}\sum_{i\in\mathcal{N}}\log\phi(B_i^{(m,t)},S_i^{(m,t)};\rho)
-\lambda|\mathcal{C}|\left(\bm{\beta}^{\top}\bm{\beta}+\bm{\nu}^{\top}\bm{\nu}\right)
+\text{const.}\nonumber\\
&:=Q_1^{(t)}(\bm{\alpha},\bm{\beta})+Q_2^{(t)}(\bm{\gamma},\bm{\nu},\psi)+Q_3^{(t)}(\rho)
\end{align}
with \(\log \left[p_k^{(m,t)}(\bm{\alpha},\bm{\beta})/(1-p_k^{(m,t)}(\bm{\alpha},\bm{\beta}))\right]:=(\bm{\alpha}^{\top},\bm{\beta}^{\top})\bar{\bm{X}}_k^{(m,t)}\) and \(\log \mu_k^{(m,t)}(\bm{\gamma},\bm{\nu}):=(\bm{\gamma}^{\top},\bm{\nu}^{\top})\bar{\bm{X}}_k^{(m,t)}\), where a tiny penalty term \(\lambda|\mathcal{C}|\left(\bm{\beta}^{\top}\bm{\beta}+\bm{\nu}^{\top}\bm{\nu}\right)\) is incorporated to avoid numerical divergence of the estimated random effect coefficients, ensuring numerical stability without significantly impacting the predictions. We select a very small \(\lambda=10^{-5}\). Note that the last equality in \eqref{eq:infer:Q} showcases that the \(Q\)-function can be linearly separable to three functions that depend only on different sets of parameters. Hence, we are able to maximize \(Q_1^{(t)}(\bm{\alpha},\bm{\beta})\), \(Q_2^{(t)}(\bm{\gamma},\bm{\nu},\psi)\) and \(Q_3^{(t)}(\rho)\) in parallel w.r.t. \((\bm{\alpha},\bm{\beta})\), \((\bm{\gamma},\bm{\nu},\psi)\) and \(\rho\) respectively using the Iteratively Reweighted Least Squares (IRLS) approach (\cite{jordan1994hierarchical}). The detailed IRLS procedures are presented in Appendix A.2. Hence, we obtain \(\bm{\Psi}^{(t)}:=(\bm{\alpha}^{(t)},\bm{\beta}^{(t)},\bm{\gamma}^{(t)},\bm{\nu}^{(t)},\psi^{(t)},\rho^{(t)})\), the updated parameters.

\textbf{Initialization and other computational details}: The SEM algorithm requires the input of initial parameters \(\bm{\Psi}^{(0)}:=(\bm{\alpha}^{(0)},\bm{\beta}^{(0)},\bm{\gamma}^{(0)},\bm{\nu}^{(0)},\psi^{(0)},\rho^{(0)})\) and initial MCMC proposals \((B_i^{(0,t)},S_i^{(0,t)},P_j^{(0,t)})\) for each iteration \(t\). We run a logistic Generalized Linear Model (GLM) on \(\{\tilde{Z}_k|\mathcal{D}^{\text{obs}}_k\}_{k\in\mathcal{C}}\), ignoring the effects of data missingness due to reporting time gap as a preliminary starting point, and set \(\bm{\alpha}^{(0)}\) as the resulting regression coefficient. Similarly, we run a Gamma GLM on \(\{\tilde{T}_k|\mathcal{D}^{\text{obs}}_k\}_{k\in\mathcal{C}:\tilde{Z}_k=1}\) to determine \(\bm{\gamma}^{(0)}\) and \(\psi^{(0)}\). \(\bm{\beta}^{(0)}\), \(\bm{\nu}^{(0)}\) and \(\rho^{(0)}\) are initialized randomly close to zero. For the inital MCMC proposals, one may choose \(B_i^{(0,1)}=0\), \(S_i^{(0,1)}=0\), \(P_j^{(0,1)}=0\), \(B_i^{(0,t)}=B_i^{(M,t-1)}\), \(S_i^{(0,t)}=S_i^{(M,t-1)}\) and \(P_j^{(0,t)}=P_j^{(M,t-1)}\) for \(i\in\mathcal{N}\), \(j\in\mathcal{P}\) and \(t>1\).

Additionally, we need to pre-specify the Monte Carlo sample sizes $M$ and $\mathcal{M}$ for the SE-step. Here, \(M\) sets the burn-in length (larger \(M\) reduces bias), while the size of \(\mathcal M\) determines the number of retained draws per observation (larger size of \(\mathcal M\) improves precision). Increasing either quantity raises computational cost. In our data application, \(M=20\) with \(\mathcal M=\{15,20\}\) is empirically sufficient: using larger values yielded no material gains but added substantial runtime. In the real data analysis, the SEM algorithm is iterated 200 times, which is empirically shown to be sufficient for convergence. The final estimated parameters, \(\hat{\bm{\Psi}}\), is determined as the average of the estimated parameters obtained in the last \(10\) iterations to reduce the estimation error caused by the randomness under the proposed stochastic approach, i.e., \(\hat{\bm{\Psi}}=\sum_{t=191}^{200}\bm{\Psi}^{(t)}/10\). 

Finally, we emphasize the scalability of the SEM procedure with MH-based stochastic sampling. Because each MH proposal and acceptance ratio only involve the local non-overlapping likelihood terms, the computational cost is linear in the total number of trade connections, i.e., $\mathcal{O}(|\mathcal{C}|)$. More advanced stochastic samplers such as Hamiltonian Monte Carlo (HMC) may improve time per sample, but each leapfrog step still has the same computational complexity of $\mathcal{O}(|\mathcal{C}|)$. Since our algorithm already scales linearly and is computationally feasible in our data, we do not further pursue refinements.

\subsection{Prediction} \label{sec:method_pred}
Our key modeling goals are to predict the claim occurrence probability for a new trade connection, which is important for insurers to determine fair premiums for future contracts, and to predict the probability that a claim will eventually be reported after the evaluation date \(\tau\) for an existing trade connection given that claims have not been reported yet, which is important for insurers to set up sufficient reserves to meet future obligations. For a new connection \(k'\in\mathcal{C}\), the posterior claim probability is
\begin{align} \label{eq:infer:pred_post}
\hat{p}^{\text{pos}}_{k'}&:=P(Z_{k'}=1|\mathcal{D}^{\text{obs}}_{k'},\bm{\hat{\Psi}},\tilde{\bm{Z}},\tilde{\bm{T}},\mathcal{D}^{\text{full}})=\int P(Z_{k'}=1|\mathcal{D}^{\text{obs}}_{k'},\mathcal{D}^{\text{lat}}_{k'};\hat{\bm{\Psi}})\times g(\mathcal{D}^{\text{lat}}_{k'}|\tilde{\bm{Z}},\tilde{\bm{T}},\mathcal{D}^{\text{full}})d\mathcal{D}^{\text{lat}}_{k'}\nonumber\\
&\approx\frac{1}{M}\sum_{m=1}^M\hat{p}_{k'}^{(m)}
:=\frac{1}{M}\sum_{m=1}^M\text{logit}^{-1}\left((\hat{\bm{\alpha}}^{\top},\hat{\bm{\beta}}^{\top})\bar{\bm{X}}_{k'}^{(m)}\right),
\end{align}
where \(\bar{\bm{X}}_{k'}^{(m)}=(1,\bm{X}^{B\top}_{k'},\bm{X}^{S\top}_{k'}, \tilde{\bm{U}}_{k'}^{\top},\bm{V}_{k'}^{\top},\tilde{B}_{k'}^{(m)},\tilde{S}_{k'}^{(m)},\tilde{P}_{k'}^{(m)})^{\top}\), and \((\tilde{B}_{k'}^{(m)},\tilde{S}_{k'}^{(m)},\tilde{P}_{k'}^{(m)})\) are the \(m\)-th sample of random effects simulated from the posterior distribution \(g(\mathcal{D}^{\text{lat}}_{k'}|\tilde{\bm{Z}},\tilde{\bm{T}},\mathcal{D}^{\text{full}})\) using the MCMC procedures detailed in Appendix A.3.

For an existing connection \(k\in\mathcal{C}\), we estimate the posterior probability of unreported claims, which is conditional to \(\tilde{Z}_{k}=0\) or else the result will become trivial, given by
\begin{align} \label{eq:infer:pred_ur}
\hat{p}_k^{\text{ur}}&:=P(Z_k=1|\tilde{Z}_k=0,\hat{\bm{\Psi}},\tilde{\bm{Z}},\tilde{\bm{T}},\mathcal{D}^{\text{full}})
=\int P(Z_{k}=1|\tilde{Z}_k=0,\mathcal{D}^{\text{obs}}_{k},\mathcal{D}^{\text{lat}}_{k};\hat{\bm{\Psi}}) g(\mathcal{D}^{\text{lat}}_{k}|\tilde{\bm{Z}},\tilde{\bm{T}},\mathcal{D}^{\text{full}})d\mathcal{D}^{\text{lat}}_{k}\nonumber\\
&\approx\frac{1}{M}\sum_{m=1}^M\frac{\hat{p}_{k}^{(m)}\left[1-F(\tau-\underline{d}_{\mathscr{P}(k)};\hat{\mu}_k^{(m)},\hat{\psi})\right]}{\hat{p}_{k}^{(m)}\left[1-F(\tau-\underline{d}_{\mathscr{P}(k)};\hat{\mu}_k^{(m)},\hat{\psi})\right]+\left(1-\hat{p}_{k}^{(m)}\right)},
\end{align}
where \(\hat{\mu}_k^{(m)}=\exp\left\{(\hat{\bm{\gamma}}^{\top},\hat{\bm{\nu}}^{\top})\bar{X}_k^{(m)}\right\}\), and \(\hat{p}_k^{(m)}\) and \(\bar{X}_k^{(m)}\) are defined like in \eqref{eq:infer:pred_post}. We select \(M=1000\) for prediction to ensure sufficiently accurate estimations in \eqref{eq:infer:pred_post} and \eqref{eq:infer:pred_ur}.

\section{Data Analysis} \label{sec:data anal}

\renewcommand{\arraystretch}{1.6}
\begin{table}[!h]
\caption{Summary of estimated parameters among various models to the TCI training data, along with the standard errors. Significant coefficients at 5\% level are bolded.}
\label{table: summary table}
\centering
\resizebox{\columnwidth}{!}{%
\begin{tabular}{lccccccccccccccccccc}
\hline
                                     & \multicolumn{4}{c}{\textbf{Model 1 (GLM w/ DC)}}                            & \textbf{} & \multicolumn{4}{c}{\textbf{Model 2 (GLM w/o DC)}}                           & \textbf{} & \multicolumn{4}{c}{\textbf{Model 3 (GLMM w/ DC)}}                           & \textbf{} & \multicolumn{4}{c}{\textbf{Model 4 (GLMM w/o DC)}}                          \\ \hline
                                     & \multicolumn{2}{c}{\textbf{Logit}}   & \multicolumn{2}{c}{\textbf{gamma}}   & \textbf{} & \multicolumn{2}{c}{\textbf{Logit}}   & \multicolumn{2}{c}{\textbf{gamma}}   & \textbf{} & \multicolumn{2}{c}{\textbf{Logit}}   & \multicolumn{2}{c}{\textbf{gamma}}   & \textbf{} & \multicolumn{2}{c}{\textbf{Logit}}   & \multicolumn{2}{c}{\textbf{gamma}}   \\ \cline{2-5} \cline{7-10} \cline{12-15} \cline{17-20} 
                                     & \textbf{Coef.} & \textbf{s.e.} & \textbf{Coef.} & \textbf{s.e.} & \textbf{} & \textbf{Coef.} & \textbf{s.e.} & \textbf{Coef.} & \textbf{s.e.} & \textbf{} & \textbf{Coef.} & \textbf{s.e.} & \textbf{Coef.} & \textbf{s.e.} & \textbf{} & \textbf{Coef.} & \textbf{s.e.} & \textbf{Coef.} & \textbf{s.e.} \\ \cline{2-5} \cline{7-10} \cline{12-15} \cline{17-20} 
(Intercept)                          & \textbf{-1.725}      & 0.214         & \textbf{6.274}       & 0.134         & \textbf{} & \textbf{-1.773}      & 0.214         & \textbf{6.248}       & 0.126         & \textbf{} & \textbf{-6.575}      & 0.495         & \textbf{7.118}       & 0.120         & \textbf{} & \textbf{-6.220}      & 0.518         & \textbf{6.874}       & 0.159         \\
TotalInsuredAmount           & -0.016               & 0.027         & \textbf{0.040}       & 0.016         & \textbf{} & 0.000                & 0.025         & \textbf{0.037}       & 0.017         & \textbf{} & 0.053                & 0.049         & \textbf{0.037}       & 0.012         & \textbf{} & 0.094                & 0.053         & 0.028                & 0.015         \\
Buyer-SpecificInsuredAmount                 & \textbf{0.360}       & 0.025         & \textbf{-0.045}      & 0.015         & \textbf{} & \textbf{0.390}       & 0.024         & \textbf{-0.042}      & 0.014         & \textbf{} & \textbf{0.651}       & 0.047         & \textbf{-0.086}      & 0.012         & \textbf{} & \textbf{0.637}       & 0.051         & \textbf{-0.073}      & 0.015         \\
AvgTurnoverRatio           & \textbf{0.549}       & 0.070         & \textbf{-0.095}      & 0.047         & \textbf{} & \textbf{0.527}       & 0.073         & \textbf{-0.093}      & 0.040         & \textbf{} & \textbf{0.470}       & 0.130         & \textbf{-0.133}      & 0.037         & \textbf{} & \textbf{0.542}       & 0.144         & \textbf{-0.132}      & 0.049         \\
Buyer-SpecificTurnoverRatio                 & \textbf{-1.560}      & 0.065         & -0.047               & 0.040         &           & \textbf{-1.612}      & 0.063         & -0.027               & 0.039         &           & \textbf{-2.267}      & 0.132         & \textbf{0.147}       & 0.033         & \textbf{} & \textbf{-2.354}      & 0.142         & \textbf{0.135}       & 0.042         \\
PolicyType\_Singlebuyer             & \textbf{0.537}       & 0.077         & -0.055               & 0.047         &           & \textbf{0.586}       & 0.074         & -0.065               & 0.047         &           & \textbf{0.909}       & 0.147         & \textbf{-0.107}      & 0.038         & \textbf{} & \textbf{1.014}       & 0.161         & \textbf{-0.129}      & 0.048         \\
SellerBizType\_SoleProp              & \textbf{0.215}       & 0.055         & -0.060               & 0.033         &           & \textbf{0.222}       & 0.052         & \textbf{-0.061}      & 0.031         & \textbf{} & \textbf{0.221}       & 0.103         & \textbf{-0.067}      & 0.029         & \textbf{} & 0.191                & 0.114         & -0.063               & 0.033         \\
SellerBizType\_Unspecified           & \textbf{0.346}       & 0.134         & -0.121               & 0.082         &           & \textbf{0.306}       & 0.134         & -0.121               & 0.086         &           & \textbf{0.765}       & 0.343         & -0.105               & 0.087         &           & \textbf{0.741}       & 0.341         & -0.111               & 0.089         \\
SellerBizType\_ACC                   & \textbf{-0.165}      & 0.072         & 0.044                & 0.047         &           & \textbf{-0.193}      & 0.072         & 0.043                & 0.042         &           & -0.257               & 0.151         & 0.039                & 0.035         &           & -0.266               & 0.165         & 0.050                & 0.043         \\
SellerBizType\_Listed                & -0.521               & 0.267         & -0.030               & 0.169         &           & \textbf{-0.559}      & 0.269         & -0.039               & 0.176         &           & -0.478               & 0.521         & -0.036               & 0.131         &           & -0.512               & 0.528         & -0.035               & 0.144         \\
BuyerBizType\_SoleProp               & \textbf{0.487}       & 0.072         & -0.025               & 0.042         &           & \textbf{0.357}       & 0.068         & -0.014               & 0.042         &           & \textbf{1.096}       & 0.193         & \textbf{-0.143}      & 0.044         & \textbf{} & \textbf{0.797}       & 0.178         & -0.070               & 0.043         \\
BuyerBizType\_Unspecified            & -0.098               & 0.108         & -0.006               & 0.070         &           & -0.066               & 0.112         & 0.000                & 0.074         &           & 0.046                & 0.248         & -0.047               & 0.066         &           & 0.067                & 0.248         & -0.025               & 0.077         \\
BuyerBizType\_ACC                    & \textbf{-0.667}      & 0.061         & 0.055                & 0.040         &           & \textbf{-0.531}      & 0.058         & 0.044                & 0.037         &           & \textbf{-1.388}      & 0.160         & \textbf{0.174}       & 0.033         & \textbf{} & \textbf{-1.026}      & 0.163         & \textbf{0.113}       & 0.038         \\
BuyerBizType\_Listed                 & \textbf{-2.160}      & 0.186         & 0.009                & 0.113         &           & \textbf{-1.805}      & 0.192         & -0.035               & 0.110         &           & \textbf{-4.874}      & 0.453         & \textbf{0.553}       & 0.094         & \textbf{} & \textbf{-3.589}      & 0.558         & \textbf{0.304}       & 0.142         \\
SellerIndustry\_Manufacturing        & -0.079               & 0.114         & -0.056               & 0.065         &           & -0.002               & 0.121         & -0.084               & 0.069         &           & -0.161               & 0.209         & -0.042               & 0.061         &           & -0.059               & 0.221         & -0.086               & 0.071         \\
SellerIndustry\_Wholesale            & 0.147                & 0.113         & -0.061               & 0.065         &           & \textbf{0.235}       & 0.117         & -0.084               & 0.068         &           & 0.325                & 0.209         & -0.081               & 0.060         &           & \textbf{0.456}       & 0.220         & -0.125               & 0.071         \\
SellerIndustry\_ProServices & -0.347               & 0.210         & 0.151                & 0.124         &           & -0.316               & 0.214         & 0.113                & 0.128         &           & -0.651               & 0.354         & 0.204                & 0.111         &           & -0.592               & 0.357         & 0.154                & 0.140         \\
BuyerIndustry\_Manufacturing         & \textbf{-0.158}      & 0.054         & \textbf{-0.107}      & 0.031         & \textbf{} & \textbf{-0.140}      & 0.051         & \textbf{-0.074}      & 0.026         & \textbf{} & \textbf{-0.314}      & 0.139         & \textbf{-0.078}      & 0.029         & \textbf{} & \textbf{-0.276}      & 0.119         & -0.043               & 0.031         \\
BuyerIndustry\_Wholesale             & \textbf{-0.332}      & 0.061         & -0.025               & 0.033         &           & \textbf{-0.259}      & 0.056         & -0.010               & 0.035         &           & \textbf{-0.710}      & 0.147         & 0.050                & 0.034         &           & \textbf{-0.619}      & 0.135         & 0.043                & 0.039         \\
BuyerIndustry\_ProServices  & \textbf{-1.110}      & 0.207         & -0.197               & 0.129         &           & \textbf{-1.125}      & 0.194         & -0.144               & 0.126         &           & \textbf{-2.085}      & 0.431         & 0.061                & 0.126         &           & \textbf{-1.981}      & 0.403         & 0.029                & 0.130         \\
SellerBusinessAge                    & \textbf{-0.059}      & 0.027         & 0.007                & 0.018         &           & \textbf{-0.055}      & 0.027         & 0.009                & 0.015         &           & -0.053               & 0.051         & 0.008                & 0.014         &           & -0.017               & 0.053         & 0.008                & 0.017         \\
BuyerBusinessAge                     & \textbf{-0.501}      & 0.025         & \textbf{-0.044}      & 0.015         & \textbf{} & \textbf{-0.488}      & 0.026         & \textbf{-0.049}      & 0.015         & \textbf{} & \textbf{-0.933}      & 0.072         & 0.025                & 0.017         &           & \textbf{-0.818}      & 0.070         & -0.001               & 0.017         \\
SellerAnnualSales\_Small             & \textbf{-0.376}      & 0.061         & -0.013               & 0.037         &           & \textbf{-0.381}      & 0.064         & -0.015               & 0.037         &           & \textbf{-0.600}      & 0.142         & 0.033                & 0.039         &           & \textbf{-0.570}      & 0.141         & 0.020                & 0.040         \\
SellerAnnualSales\_Medium            & \textbf{-0.548}      & 0.076         & -0.028               & 0.046         &           & \textbf{-0.555}      & 0.074         & -0.027               & 0.043         &           & \textbf{-0.896}      & 0.176         & 0.038                & 0.044         &           & \textbf{-0.853}      & 0.169         & 0.020                & 0.047         \\
SellerAnnualSales\_Large             & \textbf{-0.620}      & 0.090         & -0.061               & 0.061         &           & \textbf{-0.621}      & 0.091         & -0.052               & 0.055         &           & \textbf{-1.230}      & 0.214         & 0.036                & 0.050         &           & \textbf{-1.088}      & 0.212         & 0.000                & 0.057         \\
BuyerAnnualSales\_Small              & \textbf{-0.273}      & 0.068         & \textbf{0.090}       & 0.042         & \textbf{} & \textbf{-0.252}      & 0.070         & 0.076                & 0.044         &           & \textbf{-0.580}      & 0.158         & \textbf{0.121}       & 0.043         & \textbf{} & \textbf{-0.444}      & 0.155         & \textbf{0.088}       & 0.045         \\
BuyerAnnualSales\_Medium             & -0.074               & 0.070         & -0.003               & 0.045         &           & 0.100                & 0.072         & -0.065               & 0.047         &           & \textbf{-0.384}      & 0.170         & 0.014                & 0.043         &           & 0.046                & 0.162         & -0.081               & 0.046         \\
BuyerAnnualSales\_Large              & \textbf{-0.434}      & 0.085         & -0.029               & 0.054         &           & 0.049                & 0.081         & \textbf{-0.134}      & 0.052         & \textbf{} & \textbf{-0.697}      & 0.211         & 0.046                & 0.047         &           & 0.209                & 0.197         & \textbf{-0.150}      & 0.050         \\
Seller\(DC{_{O}}{^{(1)}}\)                        & 0.080                & 0.043         & 0.038                & 0.028         &           & -                    &               & -                    &               &           & 0.135                & 0.090         & 0.020                & 0.023         &           & -                    &               & -                    &               \\
Seller\(DC{_{I}}{^{(1)}}\)                         & -0.142               & 0.096         & -0.099               & 0.059         &           & -                    &               & -                    &               &           & -0.321               & 0.176         & -0.048               & 0.045         &           & -                    &               & -                    &               \\
Seller\(DC{_{OO}}{^{(2)}}\)                   & -0.006               & 0.028         & 0.001                & 0.019         &           & -                    &               & -                    &               &           & 0.025                & 0.052         & 0.001                & 0.012         &           & -                    &               & -                    &               \\
Seller\(DC{_{II}}{^{(2)}}\)                     & 0.000                & 0.054         & 0.051                & 0.034         &           & -                    &               & -                    &               &           & 0.028                & 0.096         & 0.000                & 0.026         &           & -                    &               & -                    &               \\
Seller\(DC{_{OI}}{^{(2)}}\)                    & -0.042               & 0.030         & -0.008               & 0.019         &           & -                    &               & -                    &               &           & 0.044                & 0.058         & -0.021               & 0.015         &           & -                    &               & -                    &               \\
Seller\(DC{_{IO}}{^{(2)}}\)                    & \textbf{0.112}       & 0.033         & 0.019                & 0.020         &           & -                    &               & -                    &               &           & \textbf{0.192}       & 0.065         & 0.009                & 0.016         &           & -                    &               & -                    &               \\
Buyer\(DC{_{O}}{^{(1)}}\)                         & \textbf{0.235}       & 0.073         & -0.043               & 0.045         &           & -                    &               & -                    &               &           & 0.306                & 0.160         & \textbf{-0.086}      & 0.043         & \textbf{} & -                    &               & -                    &               \\
Buyer\(DC{_{I}}{^{(1)}}\)                          & \textbf{0.811}       & 0.057         & \textbf{-0.073}      & 0.034         & \textbf{} & -                    &               & -                    &               &           & \textbf{1.986}       & 0.124         & \textbf{-0.226}      & 0.032         & \textbf{} & -                    &               & -                    &               \\
Buyer\(DC{_{OO}}{^{(2)}}\)                    & 0.082                & 0.048         & \textbf{0.064}       & 0.029         & \textbf{} & -                    &               & -                    &               &           & -0.165               & 0.113         & \textbf{0.087}       & 0.027         & \textbf{} & -                    &               & -                    &               \\
Buyer\(DC{_{II}}{^{(2)}}\)                      & \textbf{-0.182}      & 0.037         & 0.001                & 0.021         &           & -                    &               & -                    &               &           & 0.049                & 0.061         & \textbf{-0.044}      & 0.017         & \textbf{} & -                    &               & -                    &               \\
Buyer\(DC{_{OI}}{^{(2)}}\)                     & -0.054               & 0.047         & 0.039                & 0.028         &           & -                    &               & -                    &               &           & -0.025               & 0.099         & 0.034                & 0.027         &           & -                    &               & -                    &               \\
Buyer\(DC{_{IO}}{^{(2)}}\)                     & \textbf{-0.125}      & 0.025         & -0.010               & 0.014         &           & -                    &               & -                    &               &           & \textbf{-0.250}      & 0.049         & 0.003                & 0.013         &           & -                    &               & -                    &               \\ \hline
Dispersion Parameter                 & -                    &               & \textbf{0.324}       & 0.007         & \textbf{} & -                    &               & \textbf{0.327}       & 0.008         & \textbf{} & -                    &               & \textbf{0.272}       & 0.008         & \textbf{} & -                    &               & \textbf{0.297}       & 0.008         \\
Buyer Effect                         & -                    &               & -                    &               &           & -                    &               & -                    &               &           & \textbf{4.396}       & 0.155         & \textbf{-0.545}      & 0.025         & \textbf{} & \textbf{3.918}       & 0.158         & \textbf{-0.388}      & 0.033         \\
Seller Effect                        & -                    &               & -                    &               &           & -                    &               & -                    &               &           & \textbf{0.950}       & 0.083         & \textbf{-0.066}      & 0.022         & \textbf{} & \textbf{0.950}       & 0.090         & \textbf{-0.061}      & 0.028         \\
Policy Effect                        & -                    &               & -                    &               &           & -                    &               & -                    &               &           & \textbf{1.295}       & 0.095         & \textbf{-0.119}      & 0.025         & \textbf{} & \textbf{1.321}       & 0.104         & \textbf{-0.091}      & 0.033         \\
Corr (Buyer Effect, Seller Effect)   & -                    &               & -                    &               &           & -                    &               & -                    &               &           & \multicolumn{4}{c}{-0.018 \qquad 
 (0.038)}          &           & \multicolumn{4}{c}{-0.033 \qquad (0.038)}          \\
Loglik                               & \multicolumn{4}{c}{-38269.62}                                               &           & \multicolumn{4}{c}{-38529.54}                                               &           & \multicolumn{4}{c}{\textbf{-36620.79}}                                                        &           & \multicolumn{4}{c}{-36790.02}                                                        \\
AIC / BIC                               & \multicolumn{4}{c}{76701.24 / 77524.24}                                               &           & \multicolumn{4}{c}{77173.08 / 77752.23}                                               &           & \multicolumn{4}{c}{\textbf{73419.58} / \textbf{74323.87}}                                                        &           & \multicolumn{4}{c}{73710.04 / 74370.48}                                                        \\
Number of Parameters                 & \multicolumn{4}{c}{81}                                                      &           & \multicolumn{4}{c}{57}                                                      &           & \multicolumn{4}{c}{89}                                                      &           & \multicolumn{4}{c}{65}                                                      \\ \hline
\end{tabular}%
}
\end{table}
\renewcommand{\arraystretch}{1}

We apply our proposed directed-network variant of the bivariate GLMM, adjusted for unreported claims due to reporting time gaps, to our TCI dataset using the SEM algorithm outlined in \Cref{sec:method:inference}. For comparison, we also fit a GLM, a special case of the GLMM with $(\bm{\beta}, \bm{\nu}) = \mathbf{0}$. For both models, we consider two cases: one including DC variables and one excluding them. This allows us to assess whether incorporating DC variables and/or random effects significantly improves model performance.

\subsection{Estimation Results}
Table \ref{table: summary table} presents the estimated parameters for the bivariate GLM and GLMM, both with and without the inclusion of DC variables, applied to the training data. The GLMMs outperform the GLMs, as indicated by lower Akaike Information Criterion (AIC) and Bayesian Information Criterion (BIC) values. For both GLM and GLMM frameworks, models that include DC variables demonstrate better fitting performance than those without, evidenced by lower AIC and BIC. This suggests that incorporating DC variables and random effects significantly improves model performance. Additionally, the signs and statistical significances of most regression coefficients across all four models are generally consistent, indicating that the fitted models are robust across varying model specifications. In all models, more observed variables' regression coefficients are statistically significant under the logistic regression component than under the gamma regression component. This implies that buyer and seller observed risk characteristics (fixed effects) have higher predictive power in explaining their effects on claim probabilities than on reporting time gaps. Furthermore, all levels of random effects (buyer, seller, and policy) are statistically significant in the GLMMs, reflecting the prevalence of unobserved risk characteristics that cannot be fully explained by fixed effects in the GLMs.

We then investigate deeper into the regression coefficients to interpret the fitted models more thoroughly. Across all four models, several significant relationships emerge that offer valuable insights into the factors influencing claim probabilities. Firstly, claim probability increases with the insured amount against a specific buyer, suggesting that sellers perceive buyers with higher insured amounts as higher risk and hence they insure larger amounts to mitigate this perceived risk. Secondly, claim probability increases with the average turnover ratio of the entire policy but decreases with the turnover ratio against a single buyer. A higher turnover ratio with a specific buyer reflects more frequent payments within that trade connection, indicating a safer business environment. Conversely, when the policy-level average turnover ratio is high, a buyer-specific turnover ratio below this average signals a less secure trade relationship, increasing the claim probability. Sellers with single-buyer policies are more likely to file claims than those with multiple-buyer policies, potentially caused by adverse selection in TCI practices. Larger businesses like listed companies tend to have lower claim probabilities due to greater financial stability. Buyers in professional services industry exhibit the lowest claim probabilities across industries, as this industry is generally more financially stable and have a higher value-added nature. Higher business age for buyers is associated with lower claim probabilities, reflecting the notion that longer-established businesses are more stable and less likely to default.

When incorporating DC variables, additional patterns emerge. While most seller-side DC variables lack statistical significance, the in-outdegree for sellers is an exception. Sellers who are also buyers and purchase from a large number of well-connected sellers tend to have a higher probability of filing a claim. This may be due to their involvement in complex supply chains, increasing their exposure to systemic risks. In contrast, buyer-side DC measures are more compelling in explaining claim probability. The outdegree for buyers has a positive effect on claim probability. Buyers who connect to numerous customer businesses are part of extended supply chains and may face higher risks of non-payment from these customers, introducing contagion risk. The indegree for buyers also positively affects claim probability. Buyers connected to multiple sellers are perceived as risky by these sellers, who may collectively consider the buyer high-risk and thus seek TCI to insure against potential defaults. Furthermore, the in-outdegree for buyers is negatively related to claim probability. This suggests that a buyer's default risk tends to be lower when they purchase from larger sellers with broad networks of buyers, rather than from smaller, isolated sellers. Engaging with well-connected sellers may provide buyers with more stable supply relationships, reducing their likelihood of default.

In predicting the reporting time gap, fewer observed variables have statistically significant impacts compared to their influence on claim probabilities. Notably, reporting time lags tend to increase with the total insured amount for the entire policy but decrease with the insured amount for a single buyer. This makes sense because a seller with a larger total insured amount across the policy may have more flexibility before filing a claim for a buyer's non-payment, whereas a seller with a high insured amount for an individual buyer is more motivated to recover the substantial debt swiftly. Additionally, reporting time lags decrease with the policy's average turnover ratio; sellers with more frequent payments from buyers can detect non-payment issues more promptly and react quickly. The buyer's indegree is negatively related to the reporting time gap, as a seller can alert others about non-payment events. Therefore, when a buyer is connected to multiple sellers, any non-payment to one seller can quickly become known to others, prompting them to file claims promptly. Interestingly, the buyer's out-outdegree is positively associated with the reporting time gap. This implies that if a buyer, acting as a seller, is connected to multiple customer businesses, each with numerous downstream buyers, the seller in the trade connection may delay filing a claim for the buyer's non-payment. This delay may be due to the complexities and extended communication chains inherent in such extensive networks.

Analyzing the random effects in the bivariate GLMMs, we find that the magnitudes of the estimated coefficients for buyer-level random effects are considerably larger than those for seller and policy levels in both the logistic and gamma components. This indicates that unobserved heterogeneities among different buyers significantly contribute to the variations in claim probabilities and reporting time lags. Additionally, the coefficients of the random effects are positive in the logistic component but negative in the gamma component. This suggests that, after controlling for observed risk characteristics, more vulnerable business entities with a higher risk of non-payment are not only more likely to cause a claim but also tend to have these claims reported more promptly when a default occurs.

\subsection{Goodness-of-Fit Analysis for Reporting Time Gap}

\begin{figure}[!h]
\begin{center}
\includegraphics[width=3.5in]{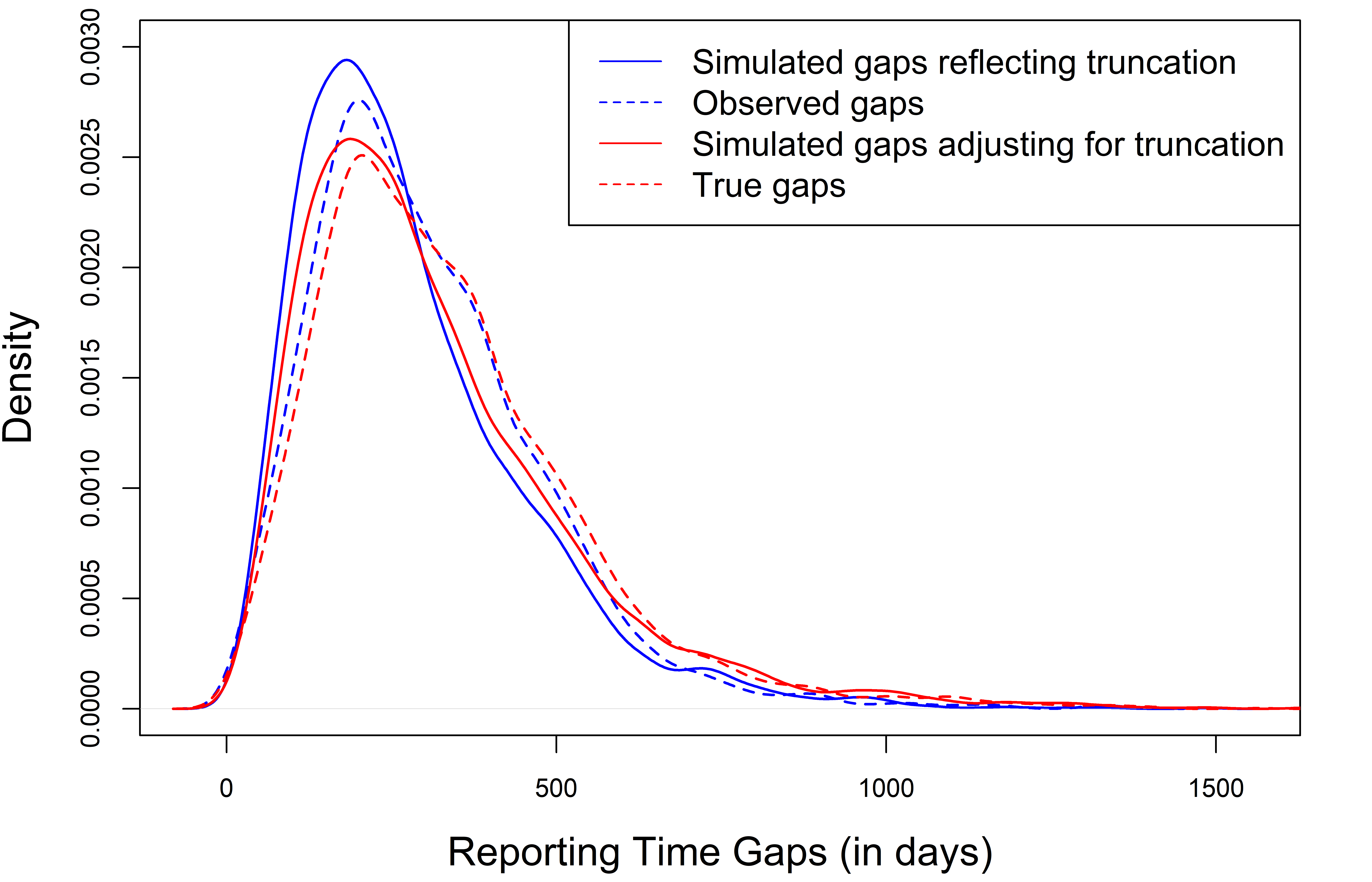}
\vspace{-0.5em}
\caption{Density plots of reporting time gaps, illustrating actual reporting time gaps (red dashed curve), observed reporting time gaps (blue dashed curve), and simulated reporting time gaps from our GLMM. The simulated reporting time gaps include those that reflect truncation (solid blue curve) and those adjusted for truncation (solid red curve). The term ``reflecting truncation'' refers to simulated reporting time gaps generated from the right-truncated fitted Gamma distribution, while ``adjusting for truncation'' refers to reporting time gaps from the complete fitted Gamma distribution.
\label{fig: Density Plots of Reporting Time Gaps}}
\end{center}
\end{figure}

\begin{figure}[!h]
\begin{center}
\includegraphics[width=0.49\textwidth]{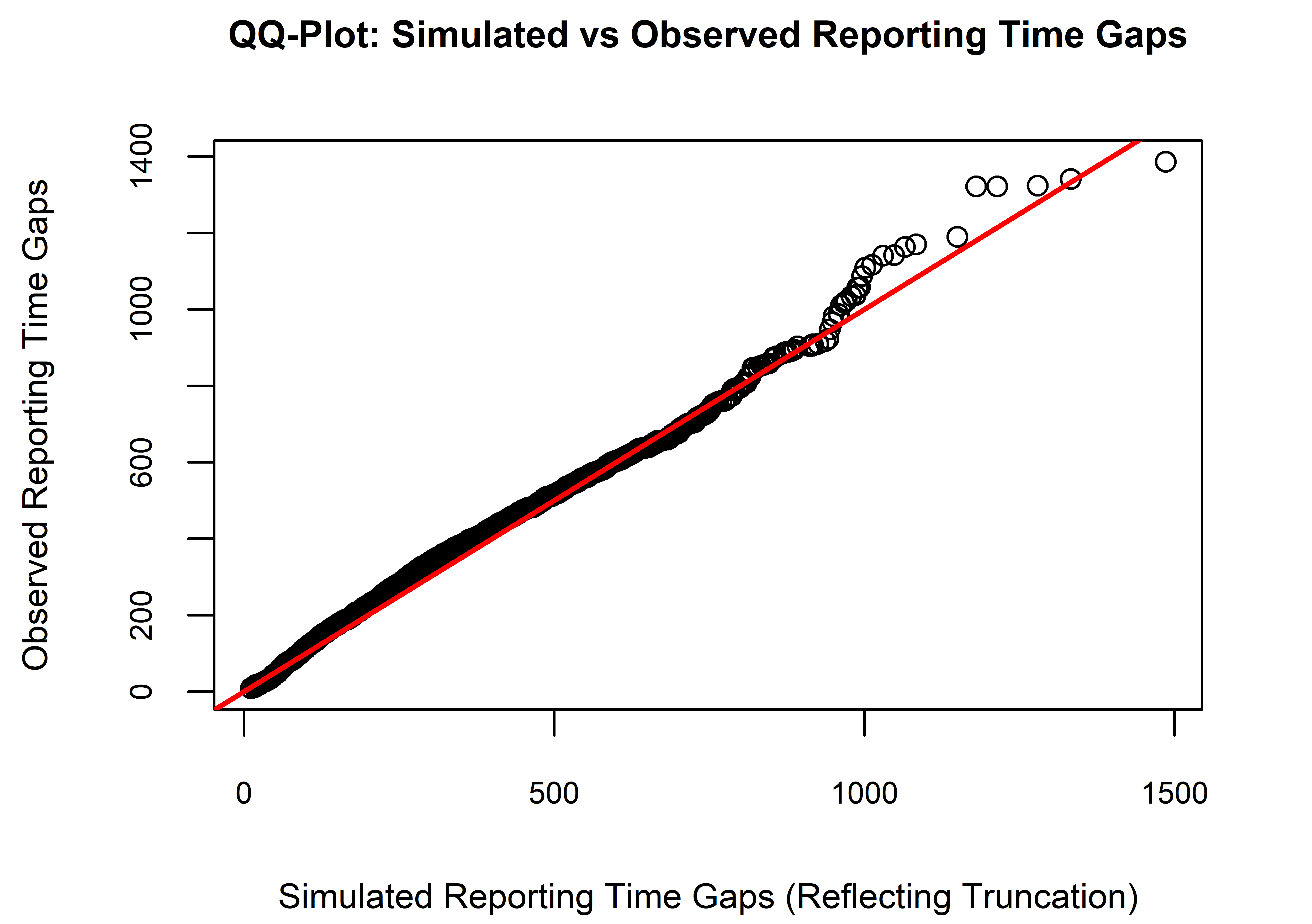}
\includegraphics[width=0.49\textwidth]{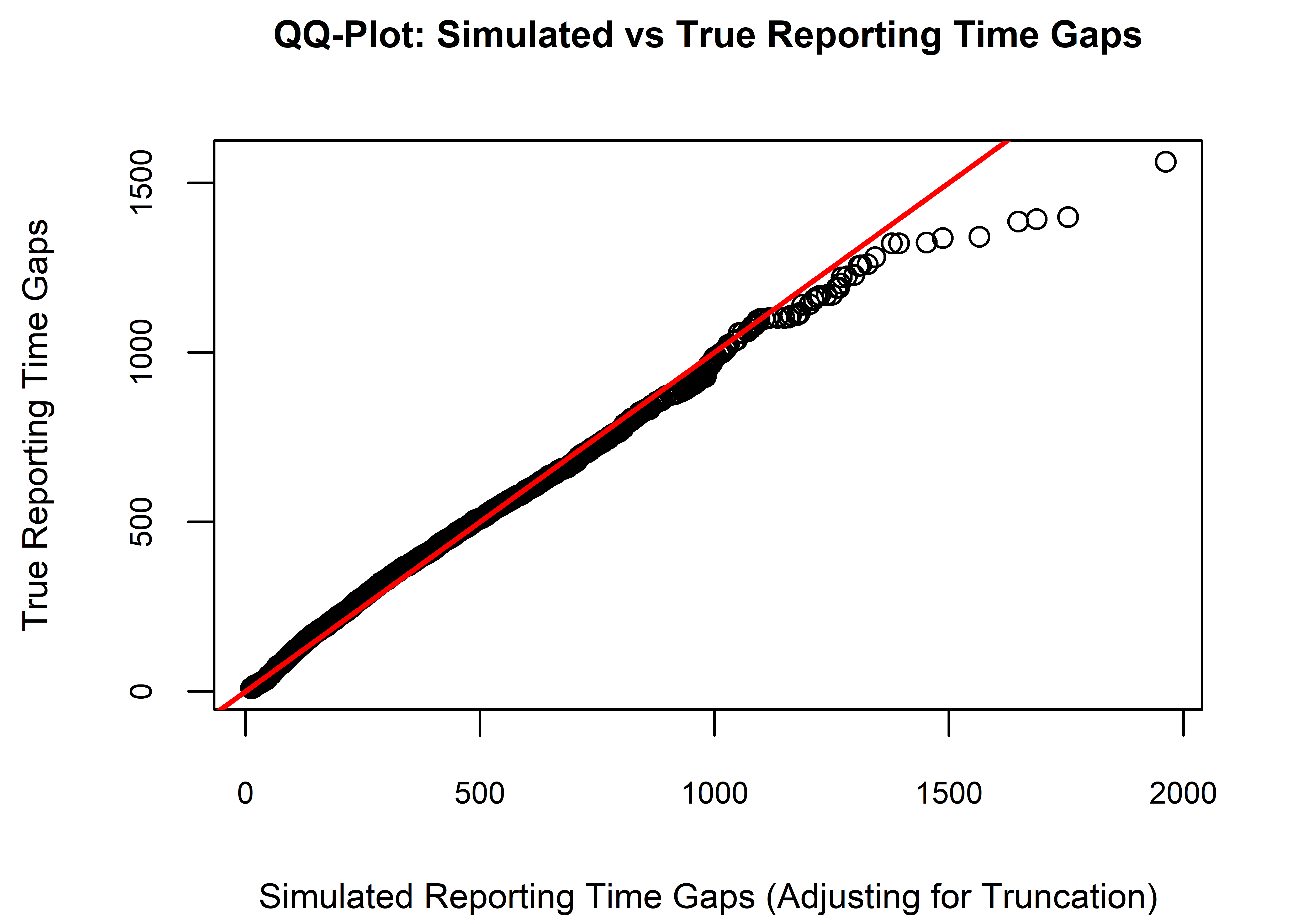}
\end{center}
\vspace{-1.5em}
\caption{Q-Q plots of fitted versus empirical reporting time gaps. Left panel: Comparing the simulated (reflecting truncation) to observed reporting time gaps. Right panel: Comparing the simulated (adjusting for truncation) to actual reporting time gaps.
\label{fig: QQ Plots: Simulated vs Observed/True Reporting Time Gaps}}
\end{figure}

As shown in Figure \ref{fig: Claims and Reporting Time Gaps}, TCI claims data are right-truncated due to unreported claims. To assess the appropriateness and importance of applying our proposed truncation adjustment method in the bivariate GLMM, we present a density plot in Figure \ref{fig: Density Plots of Reporting Time Gaps}. The alignment of the right-truncated simulated reporting time gaps with observed reporting time gaps in the training data indicates a good model fit under truncation. Furthermore, the simulated reporting time gaps that adjust for truncation align well with the full set of actual reporting time gaps, covering both observed and unreported claims, as shown in Figure \ref{fig: Claims and Reporting Time Gaps}. The noticeable rightward shift of the red curves compared to the blue curves highlights the importance of accounting for truncation. Failing to apply this adjustment may result in a substantial underestimation of reporting time gaps and increase the risk of inadequate capital reserves. We evaluate the goodness-of-fit of our GLMM for modeling the reporting time gaps using Q-Q plots in Figure \ref{fig: QQ Plots: Simulated vs Observed/True Reporting Time Gaps}, which demonstrate an excellent fit in both panels.

\subsection{Random Effects Analysis}

We illustrate how the posterior distributions of random effects, obtained using the MCMC method outlined in \Cref{sec:method_pred}, vary across claim histories and network dynamics.

\begin{figure}[!h]
\begin{center}
\includegraphics[width=5.5in]{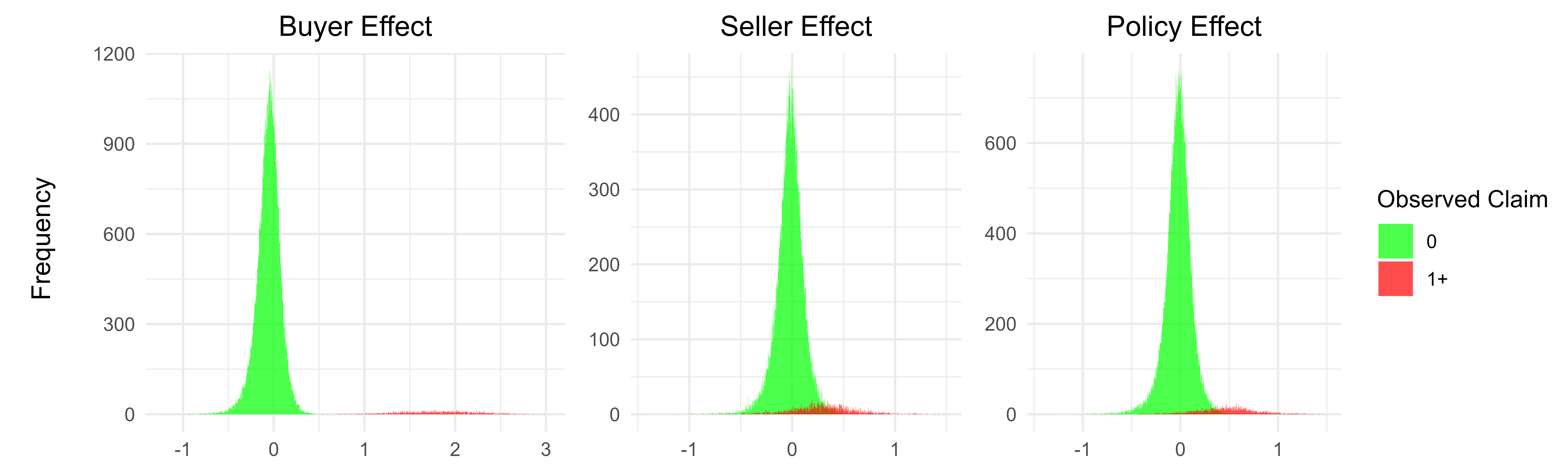}
\vspace{-0.5em}
\caption{The distribution of the posterior mean of buyer, seller, and policy-level random effects given the past claim history. Green bars represent the buyer, seller or policy with zero past observed claims, while red bars represent those with at least one past claim.
\label{fig: Empirical Distribution of Each Random Effect}}
\end{center}
\end{figure}

\Cref{fig: Empirical Distribution of Each Random Effect} presents the distributions of the posterior means of buyer, seller, and policy-level random effects under the proposed GLMM with DC. Entities without any claim history have posterior distributions highly concentrated slightly below zero for all effects, indicating they are perceived as slightly safer than those without prior information. Conversely, with at least one past claim, the posterior distributions shift substantially toward positive values and become more dispersed, suggesting that any claim history is associated with an increased likelihood of future claims. This distinction is most pronounced for the buyer effect, showing that a buyer's claim history has a greater impact on the likelihood of future claims compared to the seller's or policy's claim history. Our finding establishes a basis for differentiating risk profiles and premiums, enabling more granular ratemaking by offering lower premiums to entities without prior claims.

\begin{figure}[!h]
\begin{center}
\includegraphics[width=0.5\textwidth]{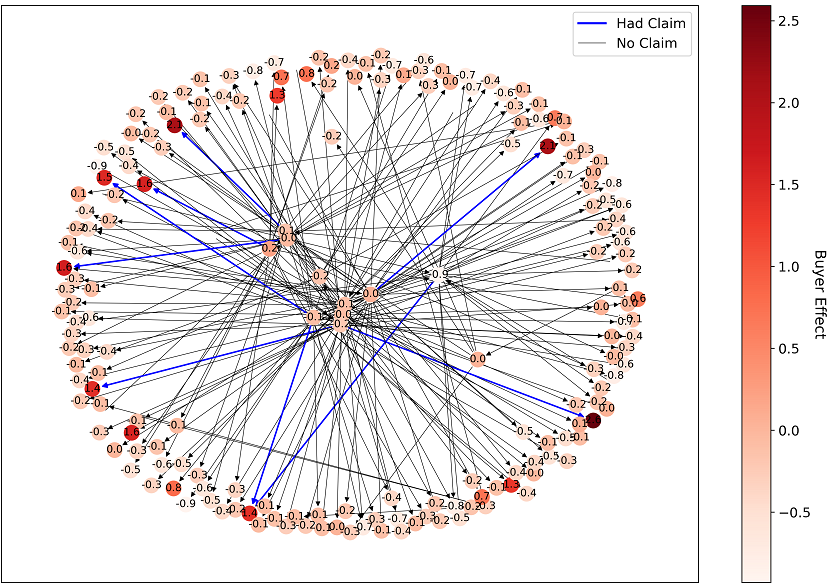}\hfill
\includegraphics[width=0.5\textwidth]{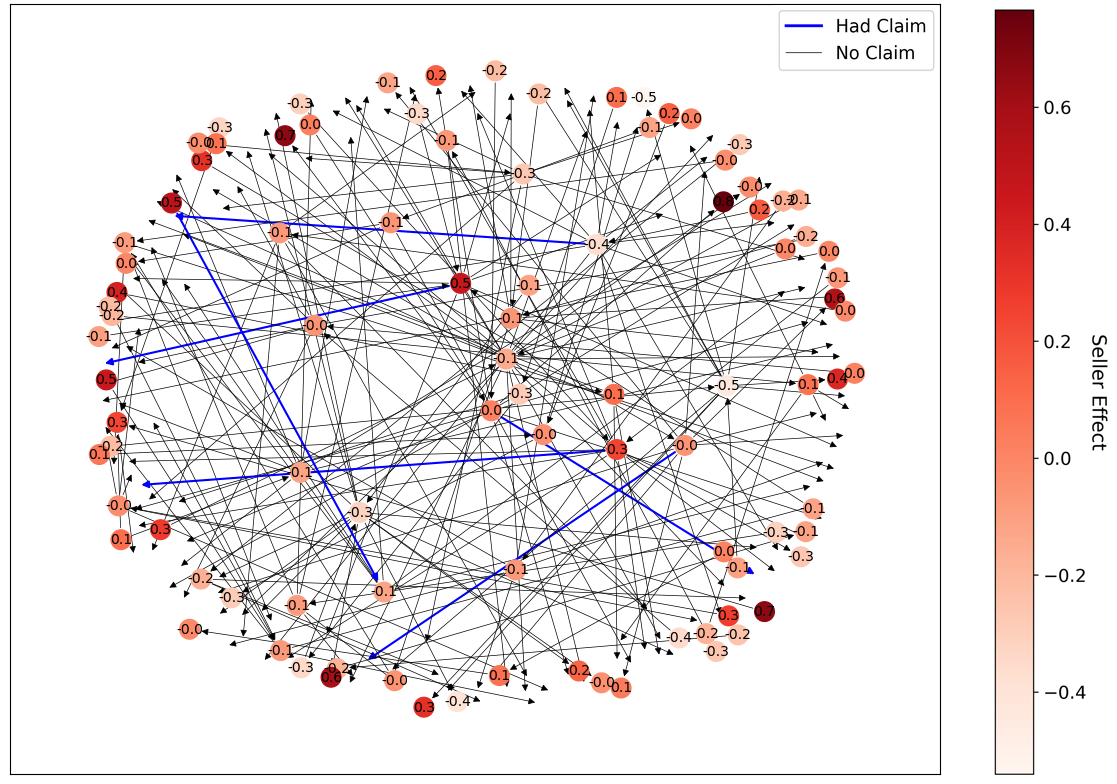}
\end{center}
\vspace{-1.5em}
\caption{A snapshot of network capturing the posterior mean of the buyer (left panel) and seller random effects (right panels). Blue arrows: Connections where a claim has occurred. Nodes with darker red color: Entities with higher posterior means of buyer or seller random effects.
\label{fig: Network Graph with Posterior Buyer and Seller Effect}}
\end{figure}

\Cref{fig: Network Graph with Posterior Buyer and Seller Effect} provides an illustrative snapshot of the network as of June 1, 2017, presenting the posterior means of the buyer and seller random effects across selected entities. In the left panel, we observe that dark red nodes are always connected by blue arrows, indicating that a buyer's past claim record is a strong signal of heightened risk for any associated future trade connections. Conversely, in the right panel, dark red nodes are not always connected to blue arrows, reflecting a weaker predictive signal of heightened risk for sellers whose past buyers have a history of non-payment.

\begin{figure}[!h]
\begin{center}
\includegraphics[width=3.5in]{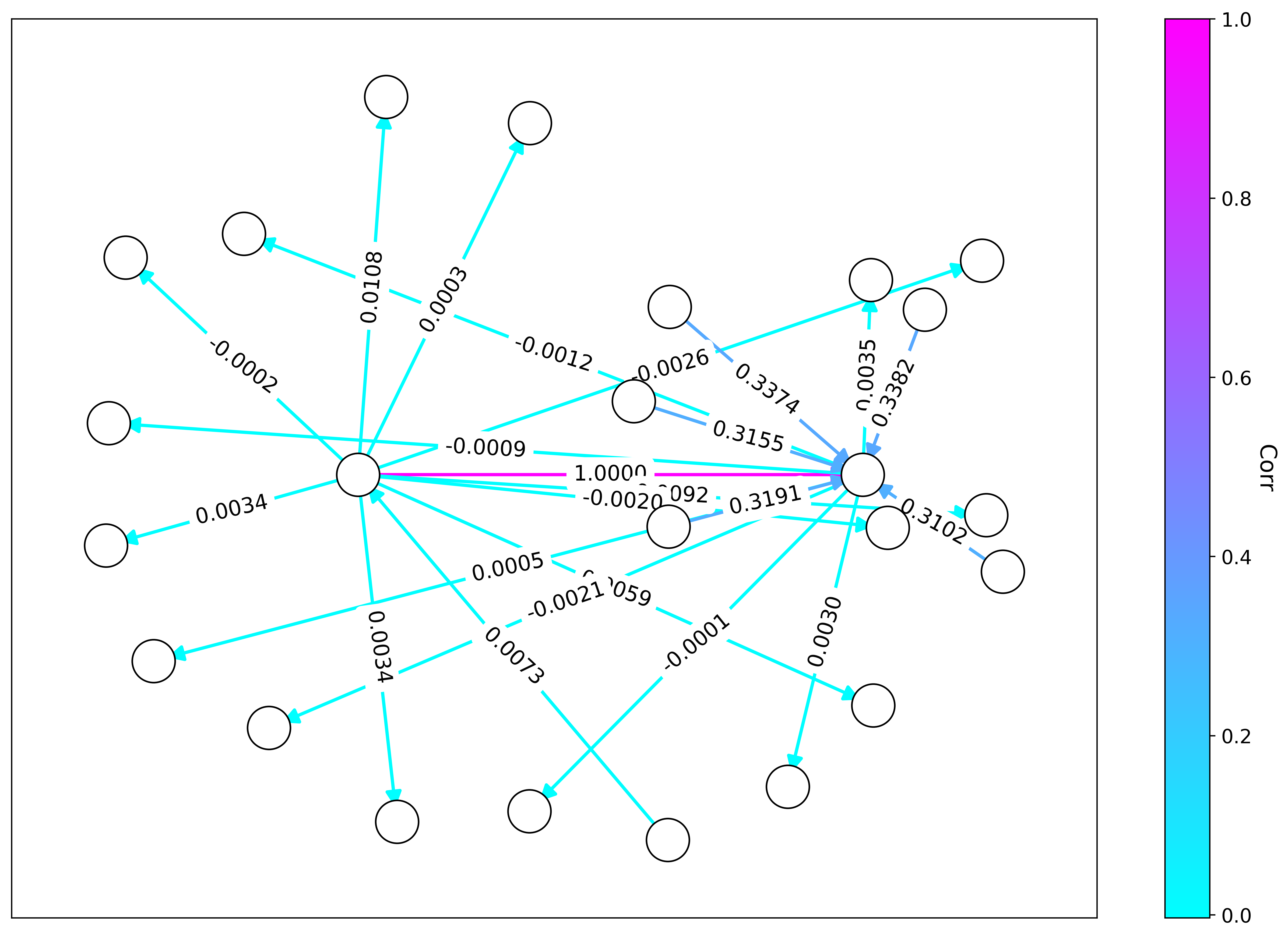}
\vspace{-0.5em}
\caption{A snapshot of network presenting the posterior conditional correlations of claim occurrence indicators between a selected trade connection (purple in color) and its adjacent connections.
\label{fig: Network Graph with Correlation between Edges}}
\end{center}
\end{figure}

\Cref{fig: Network Graph with Correlation between Edges} presents a snapshot of the posterior correlations between claim occurrence indicators for a selected connection and its adjacent connections, conditioned on the observed variables. Unlike the GLM, where all neighboring connections exhibit zero conditional correlation, the proposed GLMM reveals correlations that substantially differ from zero. We observe that only connections sharing the same buyer exhibit a high positive correlation due to the buyer effect, while other connections show minimal correlation, confirming the dominance of buyer effects over seller and policy effects. This analysis also indicates systemic risk arising from a buyer's non-payment in a buyer-seller trade connection, which can simultaneously impact other sellers connected to that buyer.

Insurers can leverage these insights to monitor relationships between entities and identify high-claim connectivity nodes, signaling contagion potential where claims propagate due to network influence rather than isolated events, enabling them to implement preventive measures to mitigate contagion risk.

\subsection{Predictive Applications}
In practice, TCI insurers rely on proprietary and confidential rating systems for pricing and reserving. However, as these mechanisms are not accessible to us, we evaluate the predictive performance of various fitted models by computing, for each trade connection in the training, validation, and test sets, the posterior probabilities of observed claims (claims that occur and are reported before the evaluation date $\tau$), unreported claims (claims that occur but are not reported until $\tau$), and complete claims (claims that eventually occur), using the methods outlined in \Cref{sec:method_pred}. We then calculate the Absolute Deviance Statistics (ADEV) for each quantity by directly comparing them to the actual outcomes, as presented in \Cref{table: ADEV}. Across all types of posterior claim probabilities and evaluation sets, GLMMs consistently exhibit lower ADEVs, indicating better fit and superior predictive performance. While performance varies slightly between GLMMs with and without DC variables, these differences are minor.

Reserves can be determined by summing the posterior unreported claim probabilities across trade connections for each training and validation set, representing the adequate capital required for the TCI company to cover unreported claims that will materialize in the future. \Cref{table: Reserves} compares the actual number of unreported claims to the estimated reserves for each model. The GLMM with DC variables aligns most closely with the actual reserves in both evaluation sets.

Since the predictive posterior claim probabilities calculated above are relevant for both pricing and reserving, our network-reinforced GLMM bridges the traditional gap between pricing and reserving (\cite{crevecoeur2023bridging}). By aligning reserving more closely with the pricing process, insurers can maintain consistency across their actuarial assumptions, improving overall portfolio management and reducing systemic discrepancies.

\begin{table}[!h]
\caption{The ADEVs for observed, unreported and complete posterior claim probabilities across various models and evaluation datasets. For the test set, ADEVs for observed and unreported claims cannot be defined, as it extends beyond the evaluation date $\tau$ of December 31, 2019. }
\label{table: ADEV}
\resizebox{\columnwidth}{!}{%
\begin{tabular}{cccccc}
\hline
\multicolumn{2}{c}{\textbf{ADEV}} & \multicolumn{1}{l}{\textbf{GLM w/DC}} & \multicolumn{1}{l}{\textbf{GLM w/o DC}} & \multicolumn{1}{l}{\textbf{GLMM w/DC}} & \multicolumn{1}{l}{\textbf{GLMM w/o DC}} \\ \hline
Training        & Observed        & 6804.60                                          & 6856.75                                           & \textbf{3125.40}                                  & 3383.34                                            \\
                & Unreported      & 1804.65                                          & 1749.87                                           & 1538.45                                           & \textbf{1401.18}                                   \\
                & Complete        & 8530.66                                          & 8536.62                                           & \textbf{4418.67}                                  & 4577.62                                            \\
Validation      & Observed        & 1704.85                                          & 1718.01                                           & \textbf{1295.34}                                  & 1314.30                                            \\
                & Unreported      & 456.98                                           & 444.09                                            & 433.47                                            & \textbf{392.54}                                    \\
                & Complete        & 2142.66                                          & 2144.80                                           & 1657.38                                           & \textbf{1650.90}                                   \\
Test         & Complete        & 2401.19                                          & 2355.61                                           & 2073.02                                           & \textbf{1913.58}                                   \\ \hline
\end{tabular}%
}
\end{table}

\begin{table}[!h]
\caption{Actual number of unreported claims versus the estimated reserves across various models and evaluation datasets.}
\label{table: Reserves}
\resizebox{\columnwidth}{!}{%
\begin{tabular}{lccccc}
\hline
\textbf{Reserve} & \textbf{Actual}      & \textbf{GLM w/DC} & \textbf{GLM w/o DC} & \textbf{GLMM w/DC} & \textbf{GLMM w/o DC} \\ \hline
\multicolumn{1}{c}{Training}    & 825                  & 1056.82                      & 993.21                        & \textbf{787.96}               & 627.68                         \\
\multicolumn{1}{c}{Validation}  & 218                  & 258.76                       & 243.87                        & \textbf{237.43}               & 191.17                         \\ \hline
                                & \multicolumn{1}{l}{} & \multicolumn{1}{l}{}         & \multicolumn{1}{l}{}          & \multicolumn{1}{l}{}          & \multicolumn{1}{l}{}          
\end{tabular}%
}
\end{table}

\section{Discussion} \label{sec:discuss}

This paper presents a novel network-augmented bivariate GLMM that incorporates entities' claim histories, detailed network relationships, and accounts for the effects of reporting time gaps. By including DC measures and multiple levels of random effects, the proposed model effectively captures the complex dependencies among entities within a network. We develop an SEM algorithm for efficient parameter estimation and demonstrate our approach using a real TCI dataset---an area previously unexplored in the literature. Our model not only provides empirical insights into the key factors affecting the riskiness of each insured trade connection but also outperforms benchmark models in terms of goodness-of-fit and predictive power. These findings highlight the importance of considering network structures to accurately predict claim probabilities for pricing and reserving purposes. 

Our empirical analysis is limited by the scope of the dataset: it contains observations from a single insurer’s portfolio and covers domestic trades only. Because our data source is the dominant domestic carrier (see Section \ref{sec:data}), we expect the observed domestic trade-connection network among entities to be broadly representative of the national domestic network structure. Additional data from other domestic insurers would likely make the network more complete and improve estimation efficiency but are not expected to fundamentally alter the substantive conclusions. If such multi-insurer data became available, our bivariate network-augmented GLMM could be extended naturally by introducing insurer-level random effects to capture cross-company heterogeneity. By contrast, analyses that involve international/export credit insurance are very different from the domestic segment in terms of, e.g., firm size, risk characteristics, and product and contractual structure. As our dataset contains domestic coverage only, we regard the analysis of international trades outside the scope of the present paper and defer it to future work contingent on data access.

Another promising direction for future research is to examine how the complete topological structure of the network influences the distributions of both claim probabilities and severities for individual trade connections. This would involve exploring alternative analytical methods beyond DC measures, which capture only limited aspects of the network's topology, to potentially enhance predictive performance. 

The proposed network-augmented bivariate GLMM methodologically targets a more general class of problems: it provides edge-level risk modeling on a directed graph with node/edge-level covariates and latent entity/policy-level effects, estimated under incomplete outcomes driven by reporting delay. Within insurance, similar data structures and modeling problems appear in many insurance products other than TCI:
\begin{itemize}
\item Cyber insurance (\cite{fahrenwaldt2018pricing}, \cite{xu2019cybersecurity}): Nodes represent firms or systems and edges represent communication or access channels. Edge-level ``attack’’ or ``compromise’’ risk reflects firm characteristics and local network configuration. Quantifying cyber-attack transmissions informs pricing and risk management. 
\item Business interruption or supply-chain insurance (\cite{rose2016improving}): Suppliers and customers form a directed network structure. Failures on upstream edges can propagate downstream, and our proposed modeling framework maps directly to edge-failure probabilities and reporting-delay components, enabling pricing and reserving in that line.
\end{itemize}
Beyond insurance, analogous edge-centric problems also arise in several important areas:
\begin{itemize}
\item Online transaction fraud (\cite{kodate2020detecting}): Buyer-seller interactions form a directed network, and fraud risk depends on user characteristics as well as the neighborhood structure, paralleling TCI’s seller-buyer directed network setting. 
\item Flight delay (\cite{sadeek2025examining}): Airports serve as nodes and flights serve as directed edges along which flight delays can propagate. Modeling edge-level delay probabilities has implications for aviation operations and flight-delay insurance.
\item Healthcare and epidemiology (\cite{simmering2015hospital}, \cite{chang2021mobility}): Hospitals or geographical locations (nodes) are connected by patient transfers or mobility flows (directed edges). The transmission risk of infectious diseases during each patient transfer or mobility flow may depend on both network structure and other observed characteristics. 
\end{itemize}

\appendix
\section*{Appendix A} \label{app:SEM}

\setcounter{subsection}{0}
\renewcommand{\thesubsection}{A.\arabic{subsection}}
\renewcommand{\theequation}{A.\arabic{equation}}
\setcounter{equation}{0}
\renewcommand{\thetable}{A.\arabic{table}}
\setcounter{table}{0}
\renewcommand{\thefigure}{A.\arabic{figure}}
\setcounter{figure}{0}

\subsection{MCMC algorithm to sample the posterior random effects in the SE-step} \label{app:SEM:effects}
The algorithm of simulating the posterior buyer, seller, and policy-level random effects involves iterating the following steps for iteration \(m=1,\ldots,M\):
\begin{enumerate}
\item Sample buyer effects. For \(i\in\mathcal{N}\):
\begin{enumerate}
    \item Propose \(\hat{B}_i^{(m,t)}=B_i^{(m-1,t)}+\widehat{\Delta B}_i^{(m,t)}\) with \(\widehat{\Delta B}_i^{(m,t)}\sim N(0,1)\).
    \item Accept \(\hat{B}_i^{(m,t)}\) with probability
    \[\min\left\{\frac{\prod_{k\in\mathcal{C}^B_i}\mathcal{L}_k^{\text{obs}}(\bm{\Psi}^{(t-1)};\tilde{Z}_k,\tilde{T}_k,\mathcal{D}_k^{\text{obs}}|\hat{B}_i^{(m,t)},S_i^{(m-1,t)},P_{\mathscr{P}(k)}^{(m-1,t)})}{\prod_{k\in\mathcal{C}^B_i}\mathcal{L}_k^{\text{obs}}(\bm{\Psi}^{(t-1)};\tilde{Z}_k,\tilde{T}_k,\mathcal{D}_k^{\text{obs}}|B_i^{(m-1,t)},S_i^{(m-1,t)},P_{\mathscr{P}(k)}^{(m-1,t)})}\frac{\phi(\hat{B}_i^{(m,t)},S_i^{(m-1,t)};\rho^{(t-1)})}{\phi(B_i^{(m-1,t)},S_i^{(m-1,t)};\rho^{(t-1)})},1\right\},\]
    where \(\mathcal{C}^B_i=\{k\in\mathcal{C}:\mathscr{B}_C(k)=i\}\).
    \item Set \(B_i^{(m,t)}=\hat{B}_i^{(m,t)}\) if the proposal is accepted, or set \(B_i^{(m,t)}=B_i^{(m-1,t)}\) if rejected.
\end{enumerate}
\item Sample seller effects. For \(i\in\mathcal{N}\):
\begin{enumerate}
    \item Propose \(\hat{S}_i^{(m,t)}=S_i^{(m-1,t)}+\widehat{\Delta S}_i^{(m,t)}\) with \(\widehat{\Delta S}_i^{(m,t)}\sim N(0,1)\).
    \item Accept \(\hat{S}_i^{(m,t)}\) with probability
    \[\min\left\{\frac{\prod_{k\in\mathcal{C}^S_i}\mathcal{L}_k^{\text{obs}}(\bm{\Psi}^{(t-1)};\tilde{Z}_k,\tilde{T}_k,\mathcal{D}_k^{\text{obs}}|B_i^{(m,t)},\hat{S}_i^{(m,t)},P_{\mathscr{P}(k)}^{(m-1,t)})}{\prod_{k\in\mathcal{C}^S_i}\mathcal{L}_k^{\text{obs}}(\bm{\Psi}^{(t-1)};\tilde{Z}_k,\tilde{T}_k,\mathcal{D}_k^{\text{obs}}|B_i^{(m,t)},S_i^{(m-1,t)},P_{\mathscr{P}(k)}^{(m-1,t)})}\frac{\phi(B_i^{(m,t)},\hat{S}_i^{(m,t)};\rho^{(t-1)})}{\phi(B_i^{(m,t)},S_i^{(m-1,t)};\rho^{(t-1)})},1\right\},\]
    where \(\mathcal{C}^S_i=\{k\in\mathcal{C}:\mathscr{S}_C(k)=i\}\).
    \item Set \(S_i^{(m,t)}=\hat{S}_i^{(m,t)}\) if the proposal is accepted, or set \(S_i^{(m,t)}=S_i^{(m-1,t)}\) if rejected.
\end{enumerate}
\item Sample policy effects. For \(j\in\mathcal{P}\):
\begin{enumerate}
    \item Propose \(\hat{P}_j^{(m,t)}=P_j^{(m-1,t)}+\widehat{\Delta P}_j^{(m,t)}\) with \(\widehat{\Delta P}_j^{(m,t)}\sim N(0,1)\).
    \item Accept \(\hat{P}_j^{(m,t)}\) with probability
    \[\min\left\{\frac{\prod_{k\in\mathcal{C}^P_j}\mathcal{L}_k^{\text{obs}}(\bm{\Psi}^{(t-1)};\tilde{Z}_k,\tilde{T}_k,\mathcal{D}^{\text{full}}|B_{\mathscr{B}_C(k)}^{(m,t)},S_{\mathscr{S}_C(k)}^{(m,t)},\hat{P}_j^{(m,t)})}{\prod_{k\in\mathcal{C}^P_j}\mathcal{L}_k^{\text{obs}}(\bm{\Psi}^{(t-1)};\tilde{Z}_k,\tilde{T}_k,\mathcal{D}^{\text{full}}|B_{\mathscr{B}_C(k)}^{(m,t)},S_{\mathscr{S}_C(k)}^{(m,t)},P_j^{(m-1,t)})}\frac{\phi(\hat{P}_j^{(m,t)})}{\phi(P_j^{(m-1,t)})},1\right\},\]
    where \(\mathcal{C}^P_j=\{k\in\mathcal{C}:\mathscr{P}(k)=j\}\).
    \item Set \(P_j^{(m,t)}=\hat{P}_j^{(m,t)}\) if the proposal is accepted, or set \(P_j^{(m,t)}=P_j^{(m-1,t)}\) if rejected.
\end{enumerate}
\end{enumerate}

Since the sets \(\{\mathcal{C}_i^B\}_{i\in\mathcal{N}}\) do not overlap, the MCMC procedure for the buyer effects can be performed in parallel for \(i\in\mathcal{N}\). Similar arguments show that parallel computing for the sampling of seller and policy effects is possible as well.

\subsection{IRLS procedures in the M-step} \label{app:SEM:IRLS}
The details regarding the IRLS procedures involved in the M-step of the proposed SEM algorithm are as follows:
\begin{enumerate}
\item Updating \((\bm{\alpha}^{(t-1)},\bm{\beta}^{(t-1)})\) to \((\bm{\alpha}^{(t)},\bm{\beta}^{(t)})\). Iterate the following until convergence:
\[(\bm{\alpha}^{\top},\bm{\beta}^{\top})^{\top}\leftarrow(\bm{\alpha}^{\top},\bm{\beta}^{\top})^{\top}-\left(\frac{\partial^2Q_1^{(t)}(\bm{\alpha},\bm{\beta})}{\partial(\bm{\alpha}^{\top},\bm{\beta}^{\top})^{\top}\partial(\bm{\alpha}^{\top},\bm{\beta}^{\top})}\right)^{-1}\frac{\partial Q_1^{(t)}(\bm{\alpha},\bm{\beta})}{\partial(\bm{\alpha}^{\top},\bm{\beta}^{\top})^{\top}},\]
where 
\[\frac{\partial Q_1^{(t)}(\bm{\alpha},\bm{\beta})}{\partial(\bm{\alpha}^{\top},\bm{\beta}^{\top})^{\top}}=\frac{1}{|\mathcal{M}|}\sum_{m\in\mathcal M}\sum_{k\in\mathcal{C}}\left(Z_k^{(m,t)}-p_k^{(m,t)}(\bm{\alpha},\bm{\beta})\right)\bar{\bm{X}}_k^{(m,t)}-2\lambda|\mathcal{C}|(\bm{0}_{\alpha}^{\top},\bm{\beta}^{\top})^{\top},\]
\begin{align*}
\frac{\partial^2Q_1^{(t)}(\bm{\alpha},\bm{\beta})}{\partial(\bm{\alpha}^{\top},\bm{\beta}^{\top})^{\top}\partial(\bm{\alpha}^{\top},\bm{\beta}^{\top})}&=-\frac{1}{|\mathcal{M}|}\sum_{m\in\mathcal M}\sum_{k\in\mathcal{C}}p_k^{(m,t)}(\bm{\alpha},\bm{\beta})\left(1-p_k^{(m,t)}(\bm{\alpha},\bm{\beta})\right)\bar{\bm{X}}_k^{(m,t)}\bar{\bm{X}}_k^{(m,t)\top}\nonumber\\
&\qquad -2\lambda|\mathcal{C}|(\bm{0}_{\alpha}^{\top},\bm{1}_{\beta}^{\top})^{\top}(\bm{0}_{\alpha}^{\top},\bm{1}_{\beta}^{\top}).
\end{align*}
Here, \(\bm{0}_{\alpha}\) and \(\bm{1}_{\beta}\) are column vectors of zeroes and ones with number of elements equals to the lengths of \(\bm{\alpha}\) and \(\bm{\beta}\) respectively.
\item Updating \((\bm{\gamma}^{(t-1)},\bm{\nu}^{(t-1)})\) to \((\bm{\gamma}^{(t)},\bm{\nu}^{(t)})\). Iterate the following until convergence:
\[(\bm{\gamma}^{\top},\bm{\nu}^{\top})^{\top}\leftarrow(\bm{\gamma}^{\top},\bm{\nu}^{\top})^{\top}-\left(\frac{\partial^2Q_2^{(t)}(\bm{\gamma},\bm{\nu},\psi^{(t-1)})}{\partial(\bm{\gamma}^{\top},\bm{\nu}^{\top})^{\top}\partial(\bm{\gamma}^{\top},\bm{\nu}^{\top})}\right)^{-1}\frac{\partial Q_2^{(t)}(\bm{\gamma},\bm{\nu},\psi^{(t-1)})}{\partial(\bm{\gamma}^{\top},\bm{\nu}^{\top})^{\top}},\]
where 
\[\frac{\partial Q_2^{(t)}(\bm{\gamma},\bm{\nu},\psi^{(t-1)})}{\partial(\bm{\gamma}^{\top},\bm{\nu}^{\top})^{\top}}=\frac{1}{\psi^{(t-1)}|\mathcal{M}|}\sum_{m\in\mathcal M}\sum_{k\in\mathcal{C}}Z_k^{(m,t)}\left(-1+\frac{T_k^{(m,t)}}{\mu_k^{(m,t)}(\bm{\gamma},\bm{\nu})}\right)\bar{\bm{X}}_k^{(m,t)}-2\lambda|\mathcal{C}|(\bm{0}_{\gamma}^{\top},\bm{\nu}^{\top})^{\top},\]
\begin{align*}
\frac{\partial^2Q_2^{(t)}(\bm{\gamma},\bm{\nu},\psi^{(t-1)})}{\partial(\bm{\gamma}^{\top},\bm{\nu}^{\top})^{\top}\partial(\bm{\gamma}^{\top},\bm{\nu}^{\top})}&=-\frac{1}{\psi^{(t-1)}|\mathcal{M}|}\sum_{m\in\mathcal M}\sum_{k\in\mathcal{C}}Z_k^{(m,t)}\bar{\bm{X}}_k^{(m,t)}\bar{\bm{X}}_k^{(m,t)\top} -2\lambda|\mathcal{C}|(\bm{0}_{\gamma}^{\top},\bm{1}_{\nu}^{\top})^{\top}(\bm{0}_{\gamma}^{\top},\bm{1}_{\nu}^{\top}).
\end{align*}
Here, \(\bm{0}_{\gamma}\) and \(\bm{1}_{\nu}\) are column vectors of zeroes and ones with number of elements equals to the lengths of \(\bm{\gamma}\) and \(\bm{\nu}\) respectively.
\item Computing \(\psi^{(t)}\). We numerically maximize \(Q_2^{(t)}(\bm{\gamma}^{(t)},\bm{\nu}^{(t)},\psi)\) w.r.t. \(\psi\) to obtain \(\psi^{(t)}\).
\item Computing \(\rho^{(t)}\). Differentiating \(Q_3^{(t)}(\rho)\) w.r.t. \(\rho\) and setting the derivative to zero shows that \(\rho^{(t)}\) satisfies the following equation of $\rho$:
\begin{equation} \label{supp:eq:rho}
-\rho^3+(1-\Sigma_{BB}^{(t)}-\Sigma_{SS}^{(t)})\rho+\Sigma_{BS}^{(t)}(\rho^2+1)=0,
\end{equation} 
where $\Sigma_{BB}^{(t)}=\frac{1}{|\mathcal{M}||\mathcal{N}|}\sum_{m\in\mathcal{M},i\in\mathcal{N}}{\left(B_i^{(m,t)}\right)}^2$, $\Sigma_{SS}^{(t)}=\frac{1}{|\mathcal{M}||\mathcal{N}|}\sum_{m\in\mathcal{M},i\in\mathcal{N}}{\left(S_i^{(m,t)}\right)}^2$, and $\Sigma_{BS}^{(t)}=\frac{1}{|\mathcal{M}||\mathcal{N}|}\sum_{m\in\mathcal{M},i\in\mathcal{N}}{B_i^{(m,t)}S_i^{(m,t)}}$.
\end{enumerate}

\begin{remark}
In our real data analysis, the quantities $\Sigma_{BB}^{(t)}$ and $\Sigma_{SS}^{(t)}$ computed in Step 4 of the algorithm above are both extremely close to 1, which is the prior variance of buyer and seller latent variables. Therefore, the solution in \eqref{supp:eq:rho} becomes $\rho^{(t)}\approx\Sigma_{BS}^{(t)}/\sqrt{\Sigma_{BB}^{(t)}\Sigma_{SS}^{(t)}}$, which is asymptotically equivalent to the empirical correlation between \(\{B_i^{(m,t)}\}_{m\in\mathcal{M},i\in\mathcal{N}}\) and \(\{S_i^{(m,t)}\}_{m\in\mathcal{M},i\in\mathcal{N}}\).
\end{remark}

\subsection{MCMC procedures to generate random effects of a new trade connection} \label{app:SEM:post}
The MCMC procedures of generating the random effects of a new trade connection \((\tilde{B}_{k'}^{(m)},\tilde S_{k'}^{(m)},\tilde P_{k'}^{(m)})\) in Section \ref{sec:method_pred} of the paper are as follows:
\begin{enumerate}
\item Simulate the posterior random effects \(\{(B_i^{(m)},S_i^{(m)},P_j^{(m)})\}_{i\in\mathcal{N},j\in\mathcal{P},m=1,\ldots,M}\) using the MCMC method with MH algorithm. Since the procedures are same as those outlined in the SE-step above except that \(\bm{\Psi}^{(t-1)}\) is replaced by \(\hat{\bm{\Psi}}\), we omit the details. 
\item Generate the posterior buyer effect samples \(\{\tilde B_{k'}^{(m)}\}_{m=1,\ldots,M}\). We consider three cases:
\begin{enumerate}
    \item If \(\mathscr{B}_C(k')\in\{\mathscr{B}_C(k):k\in\mathcal{C}\}\), i.e., the buyer from the new connection appears in the existing observed buyer poll, set \(\tilde B_{k'}^{(m)}=B_{\mathscr{B}_C(k')}^{(m)}\).
    \item If \(\mathscr{B}_C(k')\notin\{\mathscr{B}_C(k):k\in\mathcal{C}\}\) but \(\mathscr{B}_C(k')\in\{\mathscr{S}_C(k):k\in\mathcal{C}\}\), i.e., the buyer from connection \(k'\) is a new buyer that appears in the existing seller poll, simulate \(\tilde B_{k'}^{(m)}\) independently from \(N(\hat{\rho} S_{\mathscr{B}_C(k')}^{(m)},1-\hat{\rho}^2)\) for \(m=1,\ldots,M\).
    \item If \(\mathscr{B}_C(k')\notin\{\mathscr{B}_C(k):k\in\mathcal{C}\}\) and \(\mathscr{B}_C(k')\notin\{\mathscr{S}_C(k):k\in\mathcal{C}\}\), simulate \(\tilde B_{k'}^{(m)}\) iid from \(N(0,1)\) for \(m=1,\ldots,M\).
\end{enumerate}
\item Generate the posterior seller effect samples \(\{\tilde S_{k'}^{(m)}\}_{m=1,\ldots,M}\). Similar to the buyer effects outlined above, we consider three cases to determine how the seller effects are retrieved. For conciseness, we omit the details.
\item Generate the posterior policy effect samples \(\{\tilde P_{k'}^{(m)}\}_{m=1,\ldots,M}\). If \(\mathscr{P}(k')\in\mathcal{P}\), set \(\tilde P_{k'}^{(m)}=P^{(m)}_{\mathscr{P}(k')}\). Otherwise, \(\tilde P_{k'}^{(m)}\sim N(0,1)\) independently for \(m=1,\ldots,M\).
\end{enumerate}

\section*{Supplementary Materials}
The analysis code needed to prepare the TCI data, fit models, and evaluate performance with clear documentation is available at 
\href{https://github.com/tszchai/TCI}{https://github.com/tszchai/TCI}. 
A synthetic TCI dataset with 250,000 trade connections, which preserves the schema of the cleaned and de-identified real TCI data, is also provided. The synthetic data are generated entirely independently and contain no records from the real dataset, thereby avoiding any disclosure of business-sensitive information. Because the authors are bound by a non-disclosure agreement (NDA), the actual TCI data cannot be publicly released.

\end{document}